\documentclass[10pt,a4]{article}

\usepackage[utf8]{inputenc}
\usepackage{psfrag}
\usepackage[a4paper,portrait,scale=.85]{geometry}
\DeclareMathAlphabet{\mathitbf}{OML}{cmm}{b}{it}
\usepackage{fancyhdr,graphics,graphicx,dcolumn,bm,fleqn,epic,eepic,float}
\usepackage{amssymb,amsmath,multirow,rotate,color}
\definecolor{red}{rgb}{1,0,0}
\definecolor{black}{rgb}{0,0,0}
\definecolor{blue}{rgb}{0,0,1}

\usepackage{setspace}
\usepackage{array} %for vertical thick lines in tables
\usepackage{multirow} %multirow tables
\usepackage{nicefrac} %for fractions like 1/4
\def\addsymbol #1: #2#3{$#1$ \> \parbox{5in}{#2 \dotfill \pageref{#3}}\\}

\newcommand{\bite}{\begin{itemize}}
\newcommand{\eite}{\end{itemize}}
\newcommand{\benu}{\begin{enumerate}}
\newcommand{\eenu}{\end{enumerate}}

\newcommand{\beq}{\begin{equation}}
\newcommand{\eeq}{\end{equation}}
\newcommand{\beqa}{\begin{eqnarray}}
\newcommand{\eeqa}{\end{eqnarray}}
\newcommand{\barr}{\begin{array}}
\newcommand{\earr}{\end{array}}

\title{Heavy-tails in economic data: fundamental assumptions,
modelling and analysis}

\author{Jo\~ao P.~da Cruz$^{1,2}$ and Pedro G.~Lind$^{1,3}$\\
        \small{%
    $^1$ Center for Theoretical and Computational Physics, 
         University of Lisbon,}\\
        \small{%
    Av.~Prof.~Gama Pinto 2, 1649-003 Lisbon, Portugal}\\
        \small{%
    $^2$ Closer, Consultoria, Lda, Av. Eng. Duarte Pacheco Torre 2, 14C
         Lisbon, 1070-102 LISBOA, Portugal,}\\
        \small{%
    $^3$ Departamento de F\'{\i}sica, Faculdade de Ci\^encias da 
         Universidade de Lisboa, 1649-003 Lisboa, Portugal}\\
        \small{%
}
}
\begin{document}

\maketitle

%%%%%%%
\section{Introduction: fundamental postulates}
\label{intro}

The study of heavy-tailed distributions in economic and financial 
systems has been widely addressed since financial time series 
has become a research subject. In the end of $19^{th}$ century, 
Louis Jean-Baptiste Bachelier\cite{Bachelier1900}, who first suggested 
that what was latter known as Brownian Motion\cite{brownianEinstein} 
could explain stock market price fluctuations, already mentions in his 
``The Theory of Speculation'' that not all fluctuations can be 
described by Gaussian stochastic forces and in 1963 a seminal paper 
of Mandelbrot\cite{Mandelbrot1963} clearly shows that a market time series 
exhibit a much higher frequency for big variations than the expected if one 
assumes a stochastic process with Gaussian noise.
 
Nevertheless, with the introduction of Ito's lemma by Black, Scholes
and Merton\cite{Merton,BlackScholes} in European options valuation, 
Gaussian noise(GN) based processes become a market standard, 
due to the imposition of a 
revolution in the derivative market that multiplied several times business 
volumes and influenced financial market practices    
to the point where regulation itself is based on the 
same assumptions\cite{basel1,basel2,basel3}.
After the eighties, several ``highly improbable'' market drops were 
observed (e.g.~the 1987 stock market drop known as ``Black Monday'' 
and on even more recent ones, already in the 21st century) that produce 
heavy losses that were unexplainable in a GN environment.  
The losses incurred in these large market drop events did not change 
significantly the market practices or the way regulation is done
but drove some attention back to the study of heavy-tails and their underlying 
mechanisms. Some recent findings in these context is the scope of this
chapter.

Heavy-tails of financial and economic variables distribution can be 
investigated through two different approaches.

One is what we call the epistemological approach, in which the behavior 
of the economic system is explained by a few key features of the 
behavior itself, e.g.~the amplitude of price fluctuations or the 
analytical form of the return 
distributions\cite{Mantegna_Stanley1995,Borland_Bouchaud2005}. 
The ultimate goal in such an epistemological approach is to
construct a function from the scratch that fits any set of 
empirical data just by building up parameters until the plotted 
function fits it. 

The other approach we call the ontological approach and is based 
in what is commonly known as agent modeling\cite{Lux_report} to try to
find the mechanisms responsible for the emergence of the heavy-tails.
Agent-based models for describing and
addressing the evolution of markets has become an issue of
increasing interest\cite{abmfarmer} and has been subjected to
significant developments\cite{HaldaneMay,JohnsonLux,LuxStaufferReview}.

Still, from a point of view of market acceptance and practice adoption, 
the epistemological approach is the dominant one and both
researchers and practitioners take Gaussian distributions as {\it Ansatzs} 
for the return distributions observed in markets or,
if motivated by the findings of Mandelbrot\cite{Mandelbrot1963}, 
$\alpha$-stable L\'evy distributions or truncated  L\' evy distributions 
\cite{Mantegna_Stanley1994}. 
And, in fact, such approach would be the best one, if economic processes were 
stationary.

Unfortunately, they are not \cite{Bouchaud,Sornette,CruzLind1} 
and there is no guarantee that today's fitting on such distribution
will be the same as tomorrow's. This means that we cannot disregard 
the underlying mechanisms in the analysis of empirical data, which 
appeals for the agent modeling approach. 

The ontological approach has three main advantages\cite{Bouchaud}.
First, the system is able in this way to be decomposed directly 
into sellers and buyers, being a straightforward translation of
the finance system itself. 
Second, one does not need to assume the system to be in equilibrium, 
a very important aspect regarding the fact that financial markets are 
not systems in equilibrium. 
Third, by properly incorporating the ingredients of financial agents 
and the trades among them one can directly investigate the impact
of trades in the price, according to some prescribe scheme, investigating
directly possible future scenarios of the finance system.

From the bottom to the top, looking from the agent model to the distribution 
that results from their aggregation,
both Gaussian and heavy-tailed distributions are of interest.
The common usage of either 
distribution types is given by the Central Limit 
Theorem\cite{Embrechts,Billingsley,Mantegna_Stanley1995}, which states roughly
that the aggregation of a growing number of random variables converges to 
an $\alpha$-stable L\'evy distribution.
More precisely, 
in its classical and stronger version, the Central Limit Theorem 
states that a sequence of 
independent and identically distributed random variables with finite 
expected values and variance, sum up to a 
Gaussian distribution when the sample size is increased.
Lyapunov version of the theorem 
proves \cite{Billingsley} that the same is valid in 
the case where the random variables that are not identically distributed, 
requiring that they have moments of order greater than two and its
growth rate limited by the Lyapunov condition or Lindeberg 
condition \cite{Embrechts,Billingsley}.

In both cases the random variables are independent from each other.
If they are not, one may still group sets of such 
dependent variables into 
independent and identical distributed clusters of variables and, 
again, the result is a Gaussian distribution.  
One exception gets out from these Gaussian distributed situations\cite{Voit}:
one where the size of these clusters is of the order of the total
size of the system, i.e.~when the second moment diverges,
$E(X^2) \rightarrow \infty$.
For such kind of dependent processes instead of a Gaussian distribution, 
a $\alpha$-stable L\'evy distribution with $\alpha < 2$ 
will result and the typical heavy tails and power-laws are observed. 

The particular case $\alpha=2$ in a stable L\'evy distribution
yields the Gaussian distribution. One feature of these $\alpha$-stable L\'evy 
distribution are their heavy tails for $\alpha<2$, which will be addressed in the following
Sections.

So, in order to a aggregation of an agent related metric 
have an heavy-tailed distribution a dependence mechanism is needed that 
bind each individual agent distribution. 
Several agent models in finance were already proposed.
The Solomon-Levy model \cite{Solomon_Levy} defines each agent as a 
wealth function $\omega_i(t)$ that cannot go bellow a floor level.
This floor level is given by $ \omega_{i}(t) \geq \omega_{0}\bar{\omega}(t)$ 
where $\bar{\omega(t)}$ is the agent average $\omega$ at instant $t$,
$\omega_0$ being a proper constant. The imposition of the floor based on the 
mean field $\bar{\omega(t)}$ 
introduces the individual distribution dependence and, on average, 
$\langle \left| w_i(t) - \bar{\omega}(t) \right|\rangle 
\sim N $ and therefore the variance scales with $N^2$. 

Other authors like Cont-Bouchaud \cite{Cont_Bouchaud} or 
Solomon-Weisbuch \cite{Solomon_Weisbush}, use percolation based models, 
which also bring up variations of system size order of magnitude and, 
obviously, will lead also to L\'evy-type distributions. 

In a nut shell, if a multi-agent model introduces a dependence mechanism that 
leads to infinite variance, it results in a heavy-tailed distribution. 
But, unlike Gaussian distribution, which can be the result of several 
different system configurations, the form of heavy-tailed distributions 
depend strongly on each detail of the system. 
Any small change in model parameters will produce different results, 
i.e.~different tail exponents. 
This means that, if Gaussian models can be safely fitted against 
empirical data, one cannot fit an heavy-tailed model to empirical data 
unless the data is produced by a system with the same model parameters, 
namely the ones that form the dependence mechanism. 
So if a multi-agent model does not give additional information and
only provides  a system with infinite second moment, there is no 
particular surplus in the application of such a model compared with a direct 
fitting of an $\alpha$-stable L\'evy distribution. Moreover, all complexity 
used to build in the individual behavior of the agents \cite{Lux_Marchesi}, 
like dividing agents by their rationality, will not contribute to the actual 
solution of the problem because is equivalent to a direct fitting.

Following the above considerations, we can eliminate from our reasoning 
every microscopic scenario that result on either weak or strong forms of 
the Central Limit Theorem that will result on a Gaussian distribution. 
What remains? Every single microscopic scenario that generates a system 
size order variance.
From this point forward we can choose arbitrarily our model composition, 
which opens a rather broad panoply of possibilities.
This mathematical freedom to choose the model for the underlying mechanisms 
is only constraint by the requirement of bridging from the biological 
foundations of an economic system to the behavior of a market curve. 
With such bridging we are able to anchor information from the market curve 
to established theoretical principles of economy without additional 
assumptions like in  \cite{CruzLind2}. We will show here that it can 
be accomplished by using epistemological and ontological approaches 
at the same time.

Here, we aim to propose a minimal agent model based in the fundamental
properties for observing heavy-tailed distributions.
In other words, 
the question we are interested in is what is the minimum model for an 
economic system, where the emergence of heavy-tails take place? 

In short, our answer is that there are three fundamental postulates for 
observing heavy-tails and therefore critical behavior and crisis in 
economic systems:

\begin{itemize}

\item[P1]
First, agents tend to deal with each other and promote trading. 
Human beings are more efficient doing 
specialized labor than being self-sufficient and for that they need to 
exchange labor\cite{Lipsey}. The more labor exchanges they make the 
more specialized they can be. 
The usage of the expression `labor' can be regarded as excessive by 
economists, but readers can look at it as the fundamental quantity that 
is common to labor, money, wage, etc. that justifies the exchange. 
Something must be common to all these quantities; if not, we wouldn't 
exchange them. The physicists can regard such fundamental quantity as 
an `economic energy'.

\item[P2]
Second, 
each agent has some attractiveness. 
Human agents are different and attract differently other 
agents to trade. As we see bellow, this difference
should reflect some imitation, where
agents tend to prefer to consume (resp.~produce) from (resp.~for) 
the agents with the largest number of consumers (resp.~producers).
The number of producer and consumer neighbors
reflects supply and demand of its labor, respectively, 
and combining both kind of neighbors should suffice to 
quantify the price of the labor exchanged.
Further, the attractiveness of each agent leads to the formation of 
an attractive field that generates a given topology, as explained below.

\item[P3]
Third, 
it is not possible for the system to consume infinite energy 
and, for that reason, a limit leverage for each agent must exist.
We only consume and produce a finite amount of the overall 
product. Thus, if an agent transposes that finite amount it loses
his (consumption) trade connectivity in the economic environment in 
order to return to an admissible state that guarantee the finiteness 
of the overall product within the system. This assumption 
is similar to the non-linear threshold of leverage introduced
by Merton\cite{Merton} in his approach to the valuation of corporate 
debt.

\end{itemize}

In this paper we show that
these three postulates P1, P2 and P3
form the minimal model that can explain 
the phenomena associated with critical behavior of the underlying
economic systems as we will show in the next sections.
In particular, heavy-tailed return distributions or power-laws emerge due to 
the economic organization and (complex) structure of trades among agents
governed by the above three postulates.

We also show that the power-law tails are characterized by an exponent
that can be measured and it is constrained by upper and lower bounds, which
we deduce analytically.
The knowledge of such boundaries are of great importance for 
risk estimates: by deriving upper and lower bounds, one avoids
underestimates which enable the occurrence of crisis unexpectedly
and avoids overestimates which prevents profit maximization of the 
trading agents.

We start in Section \ref{sec:topology} to show that postulates 
P1 and P2 naturally define a multi-agent system with an attracting
pair potential that underlies the emergence of a scale-invariant 
geometry, characterized by a heavy-tailed distribution of single
agent connections with a given exponent. 
Next, in Section \ref{sec:distributionbounderies} we
show that this geometry is constrained, namely the exponent that
describes the heavy-tails for the agent interconnections 
is bounded by minimum and maximum values. In Section 
Section \ref{modelo} we add the third postulate, P3, and translate all 
three postulates into a model that produces macroscopic observables in
economic systems. In Section \ref{finnet} we show that the model behaves
critically, transiting between two phases, and its macroscopic observables
reproduce statistically critical features observed in real economic systems.
In particular, we focus on the fluctuations of such observables
that are related to the underlying geometry of economic relations 
and, consequently, can be helpful to derived model risk measures.
In section \ref{sec:stability} we describe in detail a specific
application, showing how the model can be applied to address  the
impact of financial rules in the stability of the economic system.
Finally in Sec.~\ref{sec:conclusion} we will draw some conclusions.

%%%%%%%%%%%%%%%%%%%%%%%%%%%%%%%%%%%%%%%%%%%%%%%%%%%%%%%%%%%%%%%%%%
\section{Definition of attractive fields in economic networks}
\label{sec:topology}   

Much of advertising is about showing people that a product 
or service is chosen by many others and there is no record of a single 
case where the lack of customers is promoted as a competitive advantage.
Advertisers make use of what physicists call 
``preferential attachment''\cite{Simon,Barabasi_Albert_1999}, 
while some authors in Economics call it imitation\cite{Bouchaud}, 
and it is a representation of P1 and P2 postulates. Here we will show 
that such postulates in fact imply preferential attachment and that 
have important consequences in the topology 
of the economic system, by making use of many-particle physics.

Based on P1 and P2, we can regard a system of economic agents in the 
same way we regard a system of mechanical particles that attract each 
other.  Since every particle influences every particle, potential wells 
are formed associated with each particle in the same way orbits are formed. 
So any new particle that enters the system will choose a well were it 
will be trapped due to the dominant influence of one particle relatively 
to every others.

How can we express this dominance that makes a particle 
to choose a well and not the others? According to P2, each particle 
has a different attractiveness. Assuming that no other mechanical influence 
is present  besides the particle-particle attraction, then we can represent 
the attraction field $G_i \equiv G(M_i)$ that particle $i$ exerts over the 
other ones due to its attractiveness $M_i$. The nature of $M_i$ is irrelevant 
for our purposes, as latter will become evident.

In such a context, if a new particle enters the system, it will suffer 
the influence of all existing particles, each one with a relative intensity 
given by 
\begin{equation}
 I _i =\frac{G(M_i)}{\sum _j ^N G(M_j)}   .
 \label{eq:fraction}
\end{equation}
where $N$ is the total number of existing particles (agents), 
which we assume as $N\rightarrow \infty$. 
At first approximation, the absolute field intensity can be made equal to 
attractiveness $G(M_i)\approx M_i$ and 
\begin{equation}
 I _i =\frac{M_i}{\sum _j ^N M_j}   .
 \label{eq:fraction2}
\end{equation}
Even not knowing what is the nature of $M_i$, the relative intensity 
can be safely assumed as measurable since it is dimensionless. 
Furthermore, Eq.~(\ref{eq:fraction2}) is 
in fact the relative distribution of attractiveness over the system.
Since no other mechanical influence is present, attractiveness distribution 
can be translated as the relative number $k_i$ of neighboring agents trapped 
by the well of each agent $i$, i.e.,
\begin{equation}
 I _i\equiv P_i =\frac{k_i}{\sum_j^N k_j}   
\label{eq:fraction3}
\end{equation}
which is perfectly equivalent to a discretization of Eq.~(\ref{eq:fraction2}), 
supported as $N \rightarrow \infty$. For a better understanding, since the 
nature of attractiveness is not known, as it happens in economic systems, 
Eq. (\ref{eq:fraction3}) says that it is deduced by the number of agents 
each agent $i$ attracted, since it is an indirect measure of 
attractiveness. At the same time, statistically speaking, 
Eq.(\ref{eq:fraction3}) represents an histogram of the relative frequency 
of agents attracted by other agents and, by the Law of Large 
Numbers\cite{Embrechts}, it converges to a probability distribution 
that equals $P_i$ in Eq.(\ref{eq:fraction3}).

The expression (\ref{eq:fraction3}) has been used as an assumption on 
several previous works  
\cite{Bouchaud,Barabasi_Albert_1999,Dorogovtsev_Mendes_Samukhin_2000,Gonzalez06}
for the representation of the natural tendency that humans have to connect to 
the ones that already have more connections.
As we will deduced below, there is an important consequence of this 
conclusion\cite{jeong}:
if only this attractiveness is considered in the network, the 
probability of randomly finding a body with $k_{i}$ connections is
\begin{equation}
 \Pi(k_{i}) \propto k_{i} ^{-\gamma}
\label{eq:powerlaw}
\end{equation}
where $\gamma$ is one exponent characterizing the system topology.
A power-law as the one in Eq.~(\ref{eq:powerlaw}) is a rather general law 
for the structure of a system with different bodies out of 
equilibrium, and it can be observed in many economic-like systems ranging
from airports\cite{Li} to the World Wide Web\cite{Broder}. 

Thus, from this section we conclude that economic agents organize 
themselves in a scale invariant topology given by Eq.~(\ref{eq:powerlaw}),
which follows from postulates P1, P2 and the Law of Large Numbers.

%%%%%%%%%%%%%%%%%%%%%%%%%%%%%%%%%%%%%%%%%%%%%%%%%%%%%%%%%%%%%%%%
\section{From geometric constraints in economic networks to bounded heavy-tails}
\label{sec:distributionbounderies}   
 
Speaking about geometrical constraints in Economics may sound odd and too
abstract but the fundamental message that will be evident bellow, tells
that heavy-tails in Economy are characterized by exponents whose values
are bounded, due to the geometry of economical trading connection. 
We start by considering 
the Eq.~(\ref{eq:powerlaw}) that was derived in the previous 
section based on P1 and P2 and that has a significant meaning in terms of the 
topology of the economic system. It means that the probability 
law is scale invariant or, geometrically speaking, 
self-similar\cite{Mandelbrot}:
the entire economic system, is equal to a part of itself, at least in 
statistical terms.

The question we want to answer in this section is can $gamma$ in 
Eq.~(\ref{eq:powerlaw}) assume any (positive) value? Real economic networks 
show typically values in the 
range $2 < \gamma < 3$ \cite{barabasi,Li}, but that is taken from the 
empirical data of electronic databases for networks that have some kind 
of computational treatment. The empirical analysis of such networks
does not provide a rigorous range for every other network type, in particular
network underlying commercial and market systems. 
Next, we will show that postulates P1 and P2 imply that indeed $\gamma$ is
indeed bounded in $[2,3]$. To this end, we separate the cases of directed 
and undirected networks and use the method introduced by 
Song, Havlin and Makse \cite{Song_2005}.
%%%%%%%%%%%%%%%%%%%%%%%%%%%%%%%%%%%%%%%%%%%%%%%%%%%%%%%%%%%%%%%%%%%%%%%%%%%%%
\begin{figure*}[t]
\centerline{\includegraphics[width=0.55\linewidth]{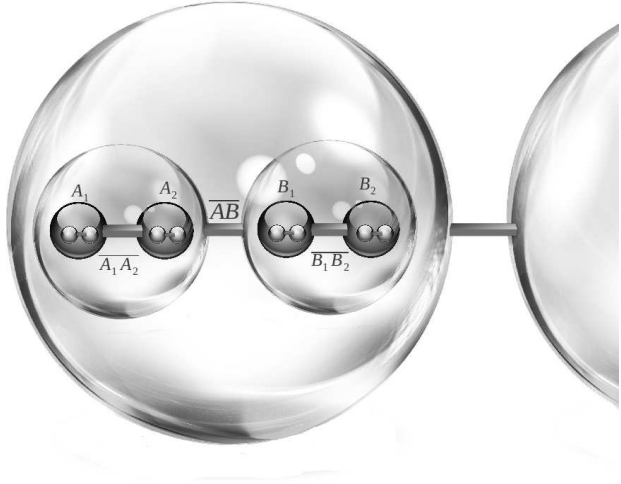}}
%\vspace*{8pt}
\caption{\protect
     Illustration of renormalization in complex networks. 
     Starting at connection $\bar{AB}$ between two clusters of agents,
     one scales down finding each cluster composed by two sets of agents,
     $A_1$ and $A_2$ on the left and $B_1$ and $B_2$ on the right connected
     again by $\bar{A_1A_2}$ and $\bar{B_1B_2}$ respectively. Each 
     sets of agents of this new ``generation'' can also be decomposed in 
     two connected sets and so on downscale.
     With such construction, in directed networks, the number of probable
     states grows with the number $Z$ of renormalized agents, since only 
     the nodes (agents) in one of the two sets can be chosen.
     Differently, for undirected networks, the number of probable states 
     grows with the square of the number of renormalized agents $Z^2$, 
     since all agents in all steps contribute to the number of probable
     states. Such difference leads to two extreme exponent values which
     bounded the possible exponents in heavy-tails of economic agent-models.}
\label{fig:directed}
\end{figure*}
%%%%%%%%%%%%%%%%%%%%%%%%%%%%%%%%%%%%%%%%%%%%%%%%%%%%%%%%%%%%%%%%%%%%%%%%%%%%%

Since the arguments for this demonstration are mainly geometric, readers 
should follow the figures together with the text. 
Figure \ref{fig:directed} represents a `renormalization' of an undirected 
network, i.e.~a network where connections have no direction, 
are symmetric: connecting $i$ to $j$ is equivalent to connection $j$ to $i$.

When renormalizing such a network, one intends at each scale to rescale
each object according to a common particular structure. 
When rescaling, connection $\bar{AB}$ gives rise to two additional connections, 
$\bar{A_1A_2}$ and $\bar{B_1B_2}$, and remains as a connection of the 
structure.
We are searching the conditions in which the distribution $P(k)$ of the
number of connections one agent has is of the form of Eq.(\ref{eq:powerlaw}).
To that end, we observe that only agents $A_1$ and $A_2$ can have their 
connectivity changed, i.e.~their degree $k$ affected, and that occurs if 
$\bar{AB}$ starts from $A_1$ or from $A_2$. So, whatever is the 
form of distribution $P(k)$ it is invariant under renormalization from one
scale (generation) to a following one if, and only if, it changes with the 
number $Z$ of new agents that emerge after renormalizing, 
i.e.~$\frac{d Z P_{d}(p,k)}{dp} = 0$ where $p$ represents the scale.

Analogously, for undirected networks, the renormalization affects the 
connectivity of all emergent agents producing this time new $Z^2$ additional 
states, since $\bar{AB}$ can connect any pair of them. 
So, in this case, $\frac{d Z^2 P_{b}(p,k)}{dp} = 0$.

Thus, we can in general write
\begin{equation}
Z^{\ell} P_d(k) dk = Z^{\ell}_{p} P_d(k_{p}) dk_{p}
\label{eq:nodedegreerenormdirec}
\end{equation}
where $\ell=1$ for directed networks and $\ell=2$ for undirected
networks, and where 
$k$ and $k_{p}$ are, respectively, an original and renormalized 
quantity of links and $Z$ and $Z_{p}$ are the correspondent quantity for 
agents. 
 
Since, according to P1 and P2, the system converges to a topology 
defined by Eq. (\ref{eq:powerlaw}), then
\begin{equation}
P(k) \rightarrow P(k _{p}) \propto k_{p} ^{-\gamma}
\label{eq:porbrenorm}
\end{equation}
If the renormalization of the links are expressed as \cite{Song_2005} 
$k \rightarrow k_{p} = \alpha_{p} k$, renormalization will result in
\begin{equation}
Z ^{\ell}  = Z_{p}^{\ell} \alpha_{p}^{-\gamma+1} .
\label{eq:nodesquare}
\end{equation}

To establish the geometrical form of the economic network, we need to 
find the relation between agent renormalization and link renormalization which 
is partially translated by Eq.(\ref{eq:powerlaw}). With 
that objective, we now define $l_{B}$ as a distance in the economic space 
between agents. The fractal dimension $d_B$ 
of the network of agents can then be calculated \cite{Song_2005} 
using a box-counting technique \cite{falconer} as
\begin{equation}
Z_{p} = Z l_{B}^{-d_{B}}
\label{eq:nodeboxcounting}
\end{equation}
and the links will scale as
\begin{equation}
k_{p} = k l_{B}^{-d_{k}} 
\label{eq:linkboxcounting}
\end{equation}
where $d_k$ is the fractal dimension of the network of links,
which yields $\alpha_{p} = l_{B}^{-d_{k}}$.

Then, using Eqs.~(\ref{eq:powerlaw}),
(\ref{eq:nodedegreerenormdirec}), (\ref{eq:nodeboxcounting}), 
and (\ref{eq:linkboxcounting}), 
the $\gamma$ value can be obtained as a function 
of the fractal dimension of both agents and links as
\begin{equation}
\gamma = 1+\ell\frac{d_{B}}{d_{k}}.
 \label{eq:gammafinal}
\end{equation}
which is the geometrical representation of $\gamma$ in a space of 
economic agents and economic links.

Some additional discussion here is needed.
Let us suppose (wrongly!) that economic networks are unweighted,
i.e.~all links are equal. 
Then, one can prove\cite{barabasi,Dorogovtsev_Mendes_Samukhin_2000,jeong}
that adding one new agent with a link to an existing agent, chosen from
probability distribution as in Eq.~(\ref{eq:fraction3}) (preferential 
attachament)\cite{Barabasi_Albert_1999}, will 
lead to a topology with degree distribution given by Eq.~(\ref{eq:powerlaw}) 
with $\gamma=3$.  That is coherent with our result (\ref{eq:gammafinal}) 
since adding one link for each agent will make links to scale like agents. 
In other words, one has $d_{B}=d_{k}$ and $\gamma=3$ because at the end of 
each link there is one agent and for each agent there is one link.
%, except for very few seed agents \cite{barabasi}.  

On the other hand, if we deny P2 and assume that all agents have equal 
attractiveness. In such a case, each node would link to every other nodes 
and we would have $Z^2-Z$ links for each set of $Z$ nodes. Meaning that,
asymptotically with $Z\rightarrow \infty$, $Z^2-Z \sim Z^2$, yielding  
$d_{k}=2d_{B}$, and $\gamma=2$.

Since economic networks are not unweighted or agents 
are not created continuously, then due to P1, links are created between 
the existing nodes leading to a weighted network. Would P2 be not
valid, links would be created randomly and, on average,  
it would result in a unweighted network, holding the above 
results. On the contrary, taking P2 as valid, we can normalize the link 
weights relatively to the maximum, meaning that $\gamma = 3$ happens 
when all weights are equal. If not, then $\gamma < 3$ because 
in such a case one has $d_{k}>d_{B}$.

In one sentence, this section leads us to the conclusion that
postulates P1 and P2 imply a bounded geometry
of the interconnected set of agents namely $2<\gamma<3$.
This result is not only corroborated by empirical data but it is
of utmost importance:
independently of the complexity associated with economic processes 
and with the network of economic relations, the economic
network is geometrically bounded. Consequently, the heavy-tailed
distributions extracted when observing the network should also
follow similar constraint which result in bounded exponent values,
as we will show below, in Section \ref{finnet}.
%%%%%%%%%%%%%%%%%%%%%%%%%%%%%%%%%%%%%%%%%%%%%%%%%%%%%%%%%%%%%%%%%%%%%%%%%%%
\begin{figure*}
\begin{center}
\includegraphics*[width=0.95\textwidth,angle=0]{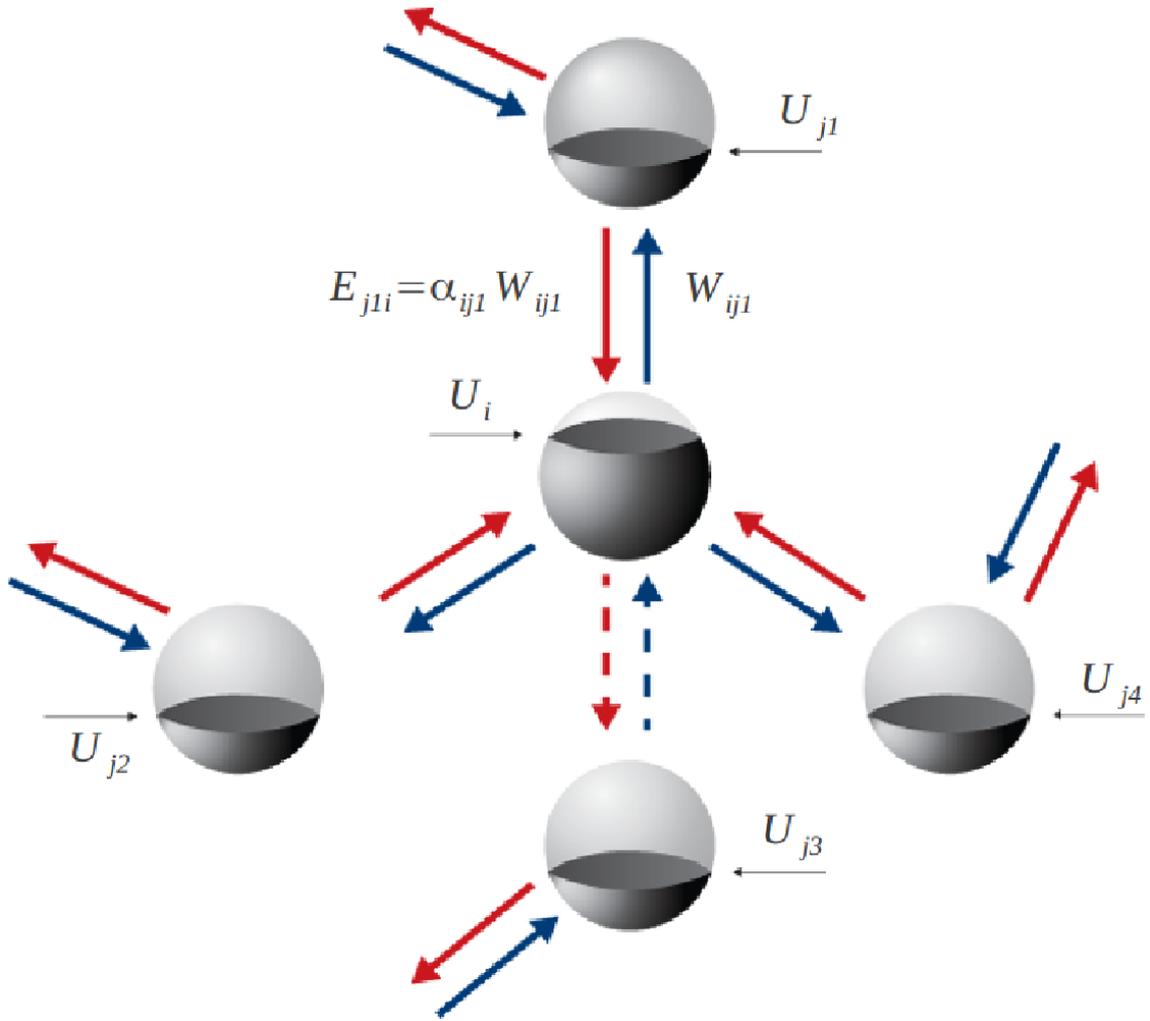}
\end{center}
\caption{\protect
  %{\bf Left:}
  Illustration of economical connections between economic agents. 
  Agent $i$ transfers an amount of ``labor'' $W_{ij_1}$ to agent $j_1$ 
  receiving in return an amount $E_{ij_1}=\alpha_{ij_1}W_{ij_1}$ where 
  $\alpha_{ij_1}$ measures how well the labor is rewarded.
  For this trade ``interaction'' agent $i$ establishes an outgoing 
  connection with (production to) agent $j_1$, while agent $j_1$ establishes 
  an incoming connection with (consumption from) agent $i$.
  The balance of this interaction yields for agent $i$ an amount
  of ``internal energy'' $U_i=\sum_{j_k} U_{ij_k}=\sum_{j_k}W_{ij_k}-E_{ij_k}$ 
  that can be summed up over 
  all agents connections. }
\label{fig1}
\end{figure*}
%%%%%%%%%%%%%%%%%%%%%%%%%%%%%%%%%%%%%%%%%%%%%%%%%%%%%%%%%%%%%%%%%%%%%%%%%%%

%%%%%
\section{Minimal agent model for the emergence of heavy-tailed 
         return distributions}
\label{modelo}

In this section we will introduce a minimal model for economical heavy-tails,
which incorporates postulates P1, P2 and, P3, introduced above in order that 
becomes possible to analyze empirical data, to simulate hypothetical
scenarios and deduce macroscopic relations.

We start by defining an economic connection as an exchange of labor 
between two agents, $i$ and $j$, which dissipates an amount 
$U_i$ in agent $i$. 
Agent $i$ delivers an amount of labor $W_{ij}$ to agent $j$ and gets a 
proportional amount of labor 
\begin{equation}
E_{ij}=\alpha_{ij} W_{ij}
\label{eij}
\end{equation} 
where $\alpha_{ij}$ is the `exchange rate' of labor. 
Figure \ref{fig1} illustrates the economic connection between one
agents and some of its neighbors $j_1$, $j_2$, $j_3$ and $j_4$.
Each one of the directed connections from $i$ to $j$, are seen 
differently by each one of the linked agents: while agent $i$ takes the 
connection as a production (or outgoing) connection, agent $j$ as a 
consumption (or incoming) connection.

Assuming that each agent can connect to several other, since every agent 
will have the propensity to establish new economic links to get 
specialized labor from other agents, the 
balance equation on each agent $i$ will be
\begin{equation}
U_{i}= \sum _{j \in L _{out} } ^{k _{i _{out}}} 
            {W _{ij}(1-\alpha _{ij})} + 
         \sum _{m \in L _{in} } ^{k _{i _{in}}} 
            {W_{mi}(\alpha _{mi} -1)} 
\label{eq: balancototal}
\end{equation}    
where  $L _{out}$ is the set of neighboring agents to which $i$ delivers 
labor and $L _{in}$ the set of neighbors from which $i$ gets labor. 
For illustrative purposes we will call henceforth $W_{ij}$, $E_{ij}$ and
$U_i$ energies, though the balance equation in Eq.~(\ref{eq: balancototal}) 
has no straightforward parallel to the energy balance in a physical
system.

Since labor in this context have arbitrary units we approximate 
\begin{equation}
W_{ij}=\langle W\rangle_{t=0}, 
\end{equation}
and representing $U_i$ in units of $\langle W\rangle$, namely
\begin{equation}
U _ {i}= k _{i,out} - k _{i,in} + \sum _{m \in L _{in} } ^{k _{i,in}} 
{\alpha _{mi}} - \sum _{j \in L _{out} } ^{k _{i,out}} {\alpha _{ij}} .
\label{eq:balancototal3}
\end{equation}   
Assuming that the exchange rate $\alpha_{ij}$ of any agent is 
approximately the rate average 
$\alpha=\left\langle \alpha _{ij} \right\rangle _ {tot}$ 
over the system, we can make a mean-field approximation as
\begin{equation}
U _ {i}= \beta ( k _{i,out} - k _{i,in}) 
\label{eq:balancototal5}
\end{equation}
with 
\begin{equation}
\beta = \left( 1-\alpha\right).
\label{beta}
\end{equation}

This balance, translated in 
Eq.~(\ref{eq:balancototal5}), between consumption and production must be 
bounded for each single agent because the only way the system dissipates 
physical energy if through the agents and dissipation must be finite 
according to the laws of thermodynamics.
Thus, agents cannot leverage themselves to infinity, i.e.~one agent cannot 
deliver more then a certain amount of labor from other agents. 
Similarly, it cannot consume an infinite amount of labor. 
This is what is reflected in postulate P3.
To account for this bounded balance that takes place for each agent, 
we define a property $c_i$ proportional to $U_i$ which is bounded
by a threshold $c_{th}$ below which the other agents break 
their production connections to it.
We can regard this threshold the same way as a limit for 
default in credit risk modeling \cite{MertonVasicek} and in the following
we derive an heuristic expression for it.

The quantity $c_i$ should not only be proportional to $U_i$ but also
inversely proportional to some fitness of the agent, namely a
quantity that measures its importance in the system, i.e.~a turnover.
The total turnover can be defined as the total number 
$k _{i,out}+k _{i,in}$ of connections involving agent $i$, both incoming and 
outgoing\cite{CruzLind1}. Also, since we are assuming P3, from a dynamical 
point of view, $k _{i,out}$ or $k _{i,in}$ can represent individually a turnover 
since, on average, the system tends to correlate both.
Thus, the threshold is defined as relative to this economic importance 
in a similar way as the usage of financial ratios as a measure of 
probability of default\cite{Crook}. By other words, 
and due to what was shown in 
the previous sections, this threshold should be invariant in connectivity 
transformations. 

For one agent $i$, considering the turnover as the total connectivity,
\begin{equation}
c_{i} = \frac{U_{i}}{T_{i}} = \frac{k_{i,out}}{k_{i,in}+}-1 .
\label{eq:defice}
\end{equation} 
Thus, each agent sees his ``relative'' internal energy $c_i$ bounded, 
namely
\begin{equation}
c_{i} \ge c_{th} 
\label{eq:threshold}
\end{equation} 
where $c_{th}$ is taken as independent from connections.
%%%%%%%%%%%%%%%%%%%%%%%%%%%%%%%%%%%%%%%%%%%%%%%%%%%%%%%%%%%%%%%%%%%%%%%%%%%
\begin{figure}[tb]
\begin{center}
\includegraphics*[width=12.0cm,angle=0]{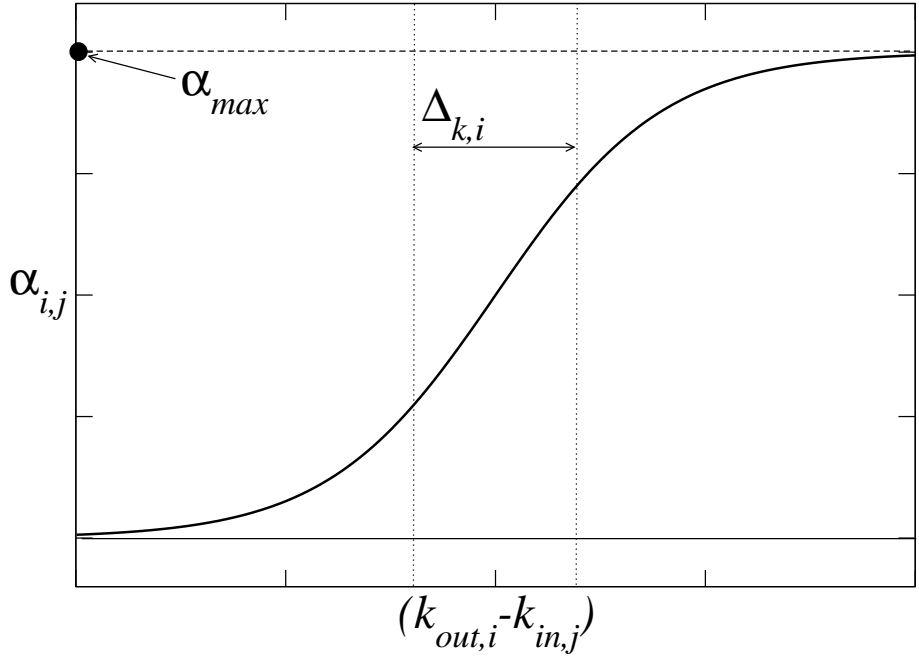}
\end{center}
\caption{\protect 
  Sketch of $\alpha_{i,j}$ in Eq.~(\ref{eq:alfaij}) as a function of 
  $k_{out,i}$ and $k_{in,j}$.
  The parameter $\Delta_{ij}$ plays the role of an absorbing length
  accounting for the sensitiveness of the system on the heterogeneity
  between trading agents (see text).}
\label{fig2}
\end{figure}
%%%%%%%%%%%%%%%%%%%%%%%%%%%%%%%%%%%%%%%%%%%%%%%%%%%%%%%%%%%%%%%%%%%%%%%%%%%

The quantity ${c}_i$ in Eq.~(\ref{eq:defice}) takes values above $-1$ 
and is well-defined for any agents with at least one incoming 
connection ($k_{i,in}>0$), a condition fulfilled by all agents,
since economic agents cannot sustained zero consumption.
The particular 
lower threshold value $\bar{c}_i=-1$ indicates an infinite leverage 
(asymptotic level), $k_{in}\to\infty$.

When Eq.~(\ref{eq:threshold}) is not fulfilled, the agent loses its 
incoming connections and will not be able to ``pay'' to its neighbors 
that will stop ``working'' for him.
When this occurs the following is imposed to the collapsed agent $i$
and its neighbors, indexed as $j$:
\begin{subequations}
\begin{eqnarray}
U _ {i} & \rightarrow & U _ {i}+\beta k_{i,in} \label{eq:avalanche1} \\
T_ {i} & \rightarrow & T_{i} - k_{i,in} \label{eq:avalanche2} \\
U _{j} & \rightarrow & U _ {j} - \beta \label{eq:avalanche3} \\
T_ {j} & \rightarrow & T_{j} - 1 \label{eq:avalanche4} .
\end{eqnarray} 
\label{eq:avalanche}
\end{subequations}

Finally, one ingredient must also be properly modeled,
namely the exchange rate defined above.
The exchange rate of labor $\alpha _{ij}$ can be taken as a dimensionless 
price and is defined heuristically as follows. See also Fig.~\ref{fig2}. 

From Eq.~(\ref{eq:balancototal5}) one can take $k _{in,i}$ as a measure
of the supply available by agent $i$. 
Similarly, $k _{out,j}$ measures the demand that one of its neighbors 
$j$ has.
When $k_{in,i}$ is large it means that many agents are working for agent
$i$ and therefore each one, in particular neighbor $j$, has a shrinked 
importance as working agent. 
Thus, $\alpha_{ij}$ should reduced with increasing $k_{in,i}$.
Similarly when $k_{out,j}$ is large it means that neighbor $j$ is
working for many other agents, each one with a proportionally small
importance, in particular the one of agent $i$. Thus, the exchange
rate for agent $i$  should increase with $k_{out,i}$.
Considering simultaneously both number of connections, $\alpha _{ij}$ 
must increase and decrease monotonically with $k_{out,j}-k_{in,i}$.
Additionally, $\alpha _{ij} \geq 0$ and the first derivatives 
must decrease when $\vert k _{out}-k _{in} \vert \rightarrow \infty$. 

Putting all such features together, $\alpha_{ij}$ must be a step function 
of $(k _{i,out} - k _{i,in})$ with $\alpha_{ij}$ attaining a maximal 
(resp.~minimal) value for $k_{out}\gg k_{in}$   (resp.~$k_{out}\ll k_{in}$).

So, our Ansatz for numeric simulation purposes is
\begin{equation}
\alpha _{ij}=\frac{\alpha_{max}}{1+e^{-\tfrac{k_{out,i}-k_{in,j}}{\Delta_{ij}}}}
\label{eq:alfaij}
\end{equation}
which is basically a step function with average value at $\alpha_{max}/2$,
similar to the hyperbolic tangent proposed by 
Cont-Bouchaud \cite{Cont_Bouchaud}. 
We call $\Delta_{ij}$ the absorbing length from agent $i$
to agent $j$.
To understand the role of the absorbing length one may easily consider
two extreme situations: (i) for $\Delta_{ij}\to \infty$ one has
$\alpha_{ij}=1$ always, i.e.~independently of the degree of agents 
$i$ and $j$ the trade is purely symmetric ($W_{ij}=E_{ij}$)
and (ii) for $\Delta_{ij}\to 0$ the dimensionless price $\alpha_{ij}$
is either $\alpha_{max}$ or $0$, depending whether $k_{out,i}>k_{in,j}$
or $k_{out,i}<k_{in,j}$ respectively and equals one for $k_{out,i}=k_{in,j}$.
Henceforth we consider $\alpha_{max}=2$ and $\Delta_{ij}=1$
without loss of generality. 

In general, other functional forms of
steps functions describing $\alpha_{ij}$ yield similar results as 
the ones shown below. 

The ``multi-agent'' model described in this section is thus a 
direct and complete translation of the three postulates P1, P2 and P3, 
the latter being introduced in order that agents are not independent 
from each other. 
Moreover, their interdependence is expressed in an intuitive form, through
Eqs.~(\ref{eq:avalanche}), associated to a nonlinear threshold 
widely use in finance and risk modeling\cite{MertonVasicek} translated
here as Eq.~(\ref{eq:defice}), completely justified by the postulates.

%%%%%%%%%%%%%%%%%%
\section{Emergence of heavy tails in empirical financial indices}
\label{finnet}

Having translated the postulates into a model, 
we next show next that under them the economic system remains at a 
critical state generating the expected heavy-tailed distributions.
To this end, we need to describe how agents chose their trading neighbors. 

We can assume that a consumption (incoming) connection 
of agent $i$ from agent $j$  occurs with a probability proportional 
to the outgoing connections (demand) agent $j$ already has.
Such scheme follows the above described imitation attachment 
mechanism\cite{Borland_Bouchaud2005}, widely study in the context of 
complex networks where it is called a preferential attachment
scheme\cite{barabasi}. 
Thus, the system of $N$ agents is initialized by addressing incoming
connections between one agents and another neighbor with a probability 
proportional to its outgoing degree. 

Every time, one new agent enters the system, e.g.~after one leaves due
to collapse (see above), the same preferential attachment scheme is
applied. See Eq.~(\ref{eq:fraction3}).

Due to the preferential attachment scheme\cite{barabasi,jeong}, 
in the initial state of the system, the outgoing
connections follow a $\delta$-distribution $P_{out}(k)=\delta(k-k_{out})$
and the incoming connections follow a scale-free distribution
$P_{in}(k)=k_{in}^{-\gamma_{in}}$,
where $\gamma_{in}$ is the exponent of the degree distribution.
As the system evolves, the number of agents remains constant
but at each event-time $n$ one new connection joining two agents
is introduced, with both agents independently chosen according 
to the preferential attachment scheme mentioned above.
Thus, through evolution both consumption and production networks
are pushed to a degree distribution of the form\cite{jeong}
\begin{equation}
P(k)\sim qk^{-\gamma}.
\label{pk}
\end{equation}
being $q$ the connection creation rate, as translation of P1.

To characterize the system during its evolution, we measure
the total system internal energy, which
accounts for all outgoing connections in the system at each time step,
namely
\begin{equation}
  U_T=\sum _{i=1}^{N}{\sum_{j\in {\cal V}_{out,i}}} {(W_{ij}-E_{ij})} ,
\label{eq:totalenergy}
\end{equation}
where ${\cal V}_{out,i}$ is the set of neighbors to which agent $i$
has outgoing connections.

We call the quantity $U_T$ the overall product and its evolution reflects 
the development or fail of the underlying economy, similar to
a financial index or GDP. Alternatively, $U_T$ can be calculated 
from the incoming connections. 
%%%%%%%%%%%%%%%%%%%%%%%%%%%%%%%%%%%%%%%%%%%%%%%%%%%%%%%%%%%%%%%%%%%%%%%%%%%
\begin{figure}[tb]
\begin{center}
\includegraphics*[width=15.0cm,angle=0]{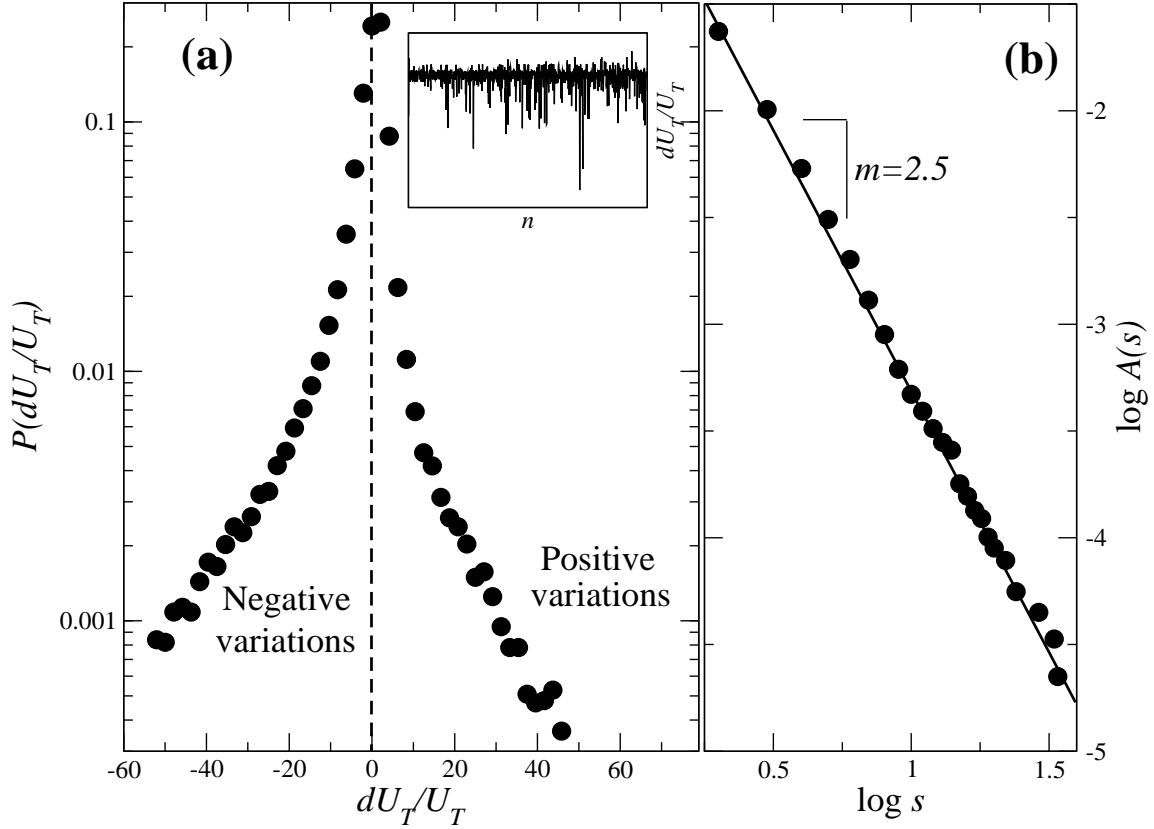}
\end{center}
\caption{\protect 
  Return distributions and avalanches size distributions in 
  the economy agent model for the variation of the total 
  internal energy, $dU_T/U_T$.
  {\bf (a)} Probability density function (PDF), showing the
  asymmetry between positive and negative variations and {\bf (b)}
  the avalanche size distribution (ASD) similar to the ones observed 
  for empirical data (compare with Fig.~\ref{fig1}).
  In the inset one sees the time series for which the PDF and the 
  ASD are computed.
  Here $N=1000$, $q=1$ and $W_{ij}=1$ for all $i$ and $j$.}
\label{fig3}
\end{figure}
%%%%%%%%%%%%%%%%%%%%%%%%%%%%%%%%%%%%%%%%%%%%%%%%%%%%%%%%%%%%%%%%%%%%%%%%%%%

Instead of the time series for $U_T$, one observes the corresponding
logarithmic returns $dU_T/U_T$, shown in the inset of Fig.~\ref{fig3}a, 
because it is a relative, dimensionless, quantity.
Figure \ref{fig3}a clearly shows that the distribution $P(dU_T/U_T)$
of the logarithmic returns is non-Gaussian. 

Parallel to $U_T$ we also keep track of the number of collapses
(see Eqs.~(\ref{eq:avalanche})) occurring successively as described
below. 
In the economical context, one such chain reaction is called
a ``crisis'' while in the physical context one calls them
``avalanches''.
Henceforth, we address to these chain reactions as crisis or 
avalanches indistinctly and we symbolize their size by $s$.
In Fig.~\ref{fig3}b the cumulative distribution $A(s)$
of the avalanche size $s$ is plotted showing a power-law whose fit yields 
$A(s)\sim s^{-m}$ with an exponent $m=2.51$ ($R^2=0.99$).

The size distribution of avalanches can be derived also in the
scope of the mean-field approach ($\alpha_{ij}\sim \alpha$)
as follows.

An agent collapses whenever Eq.~(\ref{eq:threshold}) does not hold,
which yields the following conditions:
\begin{eqnarray}
k_{i,out}- k_{i,in} &>& u _{th}(k_{i,out}+ k_{i,in})  \label{eq:eqavalanche1} \\
k_{i,out}-1- k_{i,in} &\leq& u _{th}(k_{i,out}-1+ k_{i,in}) \label{eq:eqavalanche2} 
\label{eq:eqavalanche}
\end{eqnarray} 
or more simply
\begin{eqnarray}
\omega k_{i,in} &<& k_{i,out} \leq \omega k_{i,in}+1 \label{eq:eqavalanche2_3} 
\label{eq:eqavalanche_3}
\end{eqnarray} 
with $\omega=\frac{1+u_{th}}{1-u_{th}}$. 
Therefore, the probability for agent $i$ to collapse is
\begin{equation}
P _{br} = P(\omega k_{i,in} < k_{i,out} \leq \omega k_{i,in}+1) .
\label{eq:probavalanchevizinho}
\end{equation}
Since $k_{i,in}$ and $k_{i,out}$ are integers, 
there is only one integer value for $k_{i,out}$ in the interval 
$\omega k_{i,in} < k_{i,out} \leq \omega k_{i,in}+1$,
given approximately by $k_{i,out}\sim \omega k_{i,in} + \tfrac{1}{2}$
and leading to ($\omega k_{i,in}\gg 1$):
\begin{equation}
P _{br}  \approx q(\omega k_{i,in}))^{-\gamma} .
\label{eq:probavalanchevizinho_3}
\end{equation}
%%%%%%%%%%%%%%%%%%%%%%%%%%%%%%%%%%%%%%%%%%%%%%%%%%%%%%%%%%%%%%%%%%%
\begin{figure}[H]
\centerline{\includegraphics[width=0.9\linewidth]{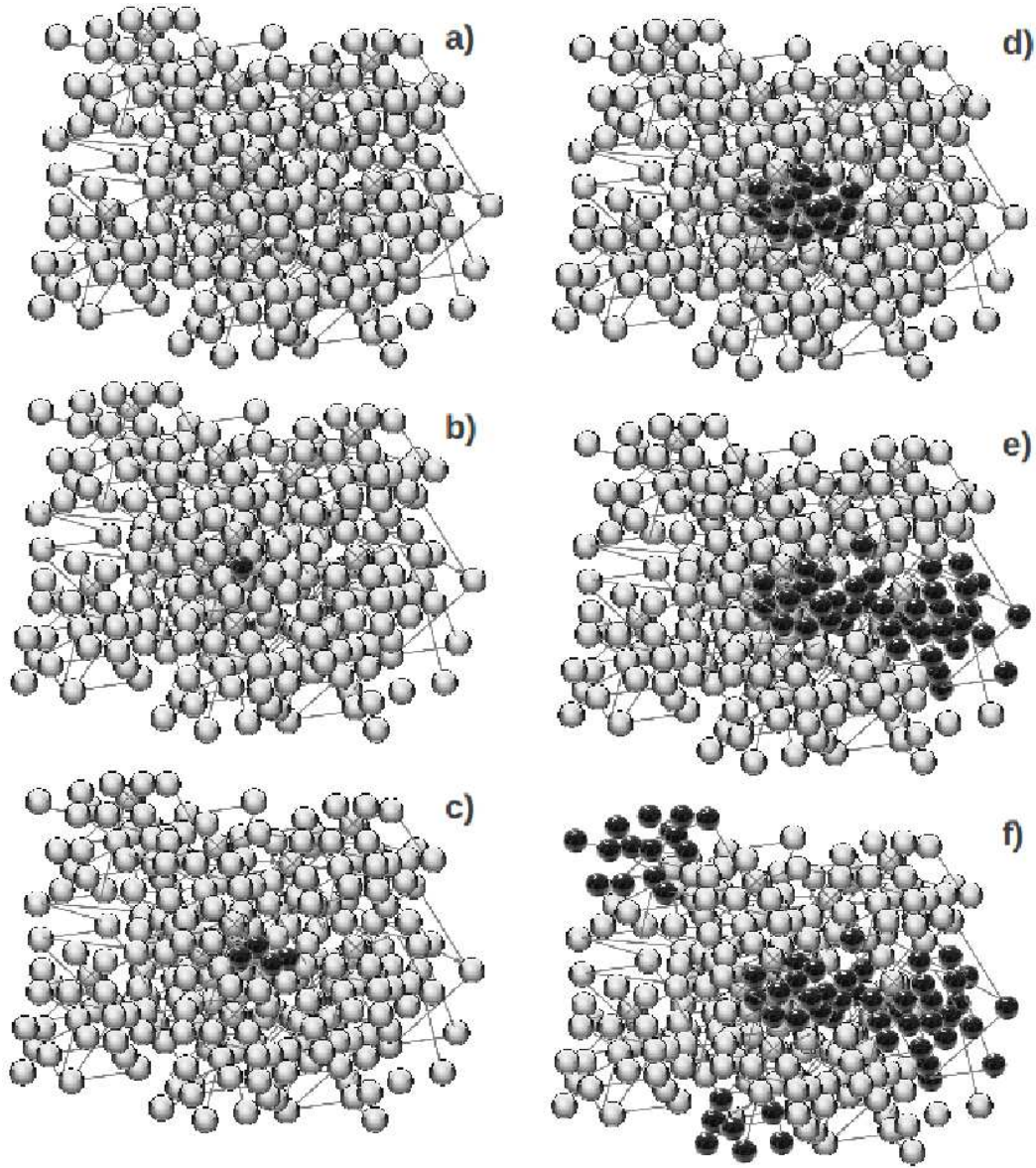}}
\vspace*{8pt}
\caption{\protect
         Illustration of a collapse chain reaction:
         {\bf (a)} The systems evolves with all $c_i\ge c_{th}$ until
         {\bf (b)} one agent collapses (black).
         The collapse of that agent may lead to {\bf (c)} 
         the collapse of some of its neighbors,
         {\bf (d)} eventually triggering a chain of collapses
         that 
         {\bf (e)} may cover a significantly region of the
         entire netwokr.
         {\bf (f)} During the evolution of the system
         independent chains of collapses (avalanches) can
         take place.}
\label{fig4}
\end{figure}
%%%%%%%%%%%%%%%%%%%%%%%%%%%%%%%%%%%%%%%%%%%%%%%%%%%%%%%%%%%%%%%%%%%

Whenever an agent collapses, if the number of neighboring agents expected
to collapse is bigger than one, the system would be in permanent destruction. 
On the other hand, if that expected number is lower than one, the system 
would grow indefinitely.
In the former case, one is lead to a scenario where agents are isolated 
and in the latter case one observes a situation of infinite energy 
consumption which is also not realistic.  
Therefore, 
the only possible state of economy is the one where the expected value of 
neighbor agents to collapse is precisely one, which corresponds to a
critical state. In practice this critical state is characterized by
a constant intermittency from one phase of economical growth
(e.g.~bubble) and another of economical drop (crisis).
Indeed critical behavior addresses precisely phenomena with such 
alternate growing and dropping regimes, as economists always said 
they exist in our modern economies and as everybody experiences
for good or worse. 

Since the expected value must be equal to one then 
\begin{equation}
\sum _{k_{i_{in}}=1} ^{\infty}{k_{i_{in}} P (k_{i_{in}})P _{br}}=1
\label{eq:valoresperadoki_0}
\end{equation}
and from Eq.~(\ref{pk})
\begin{equation}
\omega^{\gamma}= q ^{\gamma} \sum _{k_{i_{in}}=1} ^{\infty}{ (k_{i_{in}}))^{-2\gamma+1}}
              = q ^{\gamma}\zeta(2\gamma-1)
\label{eq:valoresperadoki}
\end{equation}
where $\zeta$ is Riemann Zeta function. 

With expression (\ref{eq:valoresperadoki}) we show that, only by
imposing first principles in Economy, $P1$, $P2$ and $P3$ above, 
the economic system is trapped in a critical state defined through the 
relation between economic growth $q$, leverage level $\omega$ and the 
economic organization characterized by $\gamma$. 
A similar derivation can be done for positive variations, instead of drops:
instead of limiting the leverage to collapse, one limits the savings for a 
consumption chain of events that can lead to a heavy-tailed distribution 
with different exponents on the positive and negative side since the 
mechanisms are not dependent. 

Since we know that the economical network is in a critical state,
transiting between two phases, we can look how the microscopic mechanisms 
that make neighbor agents to collapse generates the heavy-tails. Such 
mechanisms should be the same for the emergence of bubbles, but here we 
focus only in the negative part of the distribution $P(dU_T/U_T)$.

To arrive to the size distribution of avalanches further 
derivations are necessary. First one should notice that 
the collapse of a node leads to the breaking of its consumption 
links or, equivalently, its neighbors' production links. 
Meaning that each collapse will provoke a chain reaction
of size $r$, i.e., it originates a branching process as illustrated 
in Fig.~\ref{fig4} with a probability according to Otter's 
theorem \cite{Otter}, given by \cite{Otter,Harris,Athreya,lauritsen96}
\begin{equation}
P(z=r) \propto r ^{-\frac{3}{2}}
\label{eq:ottertheoremsimple}
\end{equation} 
where $z$ is the total number of nodes involved in a single branching 
process. Equation (\ref{eq:ottertheoremsimple}) holds independently of 
the topology, as long as the branching process is critical, 
but the number of collapsed agents in a real economic 
network is difficult to recount for. What is measured when an avalanche 
occurs, in such a real network, is the number of links destroyed during the avalanche.
This number of links accounts for a macroscopic property of the system,
namely the overall product $U_T$ which sums up all outgoing product 
of all agents. Therefore, we want to express $P(z)$ in terms of the 
total number of destroyed links.

From the total number $r$ of nodes included in one avalanche, since 
$P(k)\sim k^{-\gamma}$, the number of nodes with $k _j$ connections 
involved in the avalanche is $n _j=r k_j^{-\gamma}$ which
corresponds to $k_j n _j=r k_j^{-\gamma+1}$ destroyed links in 
nodes with $k_j$ links. Thus, the total number of links destroyed is 
given by
\begin{equation}
K_T=r \sum_{j=k _{min}} ^{k_{max}} {k_j^{-\gamma+1}}
\label{eq:totallinksT}
\end{equation} 
where $k_{min}$ and $k_{max}$ are the lowest and highest degree involved.
Since the sum on the right side of (\ref{eq:totallinksT}) are the degrees, 
not the links, we can substitute each $k_j$ by $\alpha_j K_T$, where 
$\alpha_j$ are leverage dependent coefficient, which leads to 
$r \propto K_T^{\gamma}$, i.e.~$P(K_T) \propto K_T^{-\frac{3}{2}\gamma}$.

So, the fraction of avalanches of size -- number of lost links --
larger than size $s$ is given by
\begin{equation}
P(K_T\geq s)\propto \int _{s} ^{+\infty} {x ^{-\frac{3}{2}\gamma} dx} 
            \propto s ^{-\frac{3}{2}\gamma+1} \equiv s^{-m}
\label{eq:avalanchefinalacum}
\end{equation}    
yielding 
\begin{equation}
m=\tfrac{3}{2}\gamma-1
\label{eq:avalanchefinalacum2}
\end{equation}
%%%%%%%%%%%%%%%%%%%%%%%%%%%%%%%%%%%%%%%%%%%%%%%%%%%%%%%%%%%%%%%%%%%%%%%%%%
\begin{figure}[H]
\centerline{\includegraphics[width=0.9\linewidth]{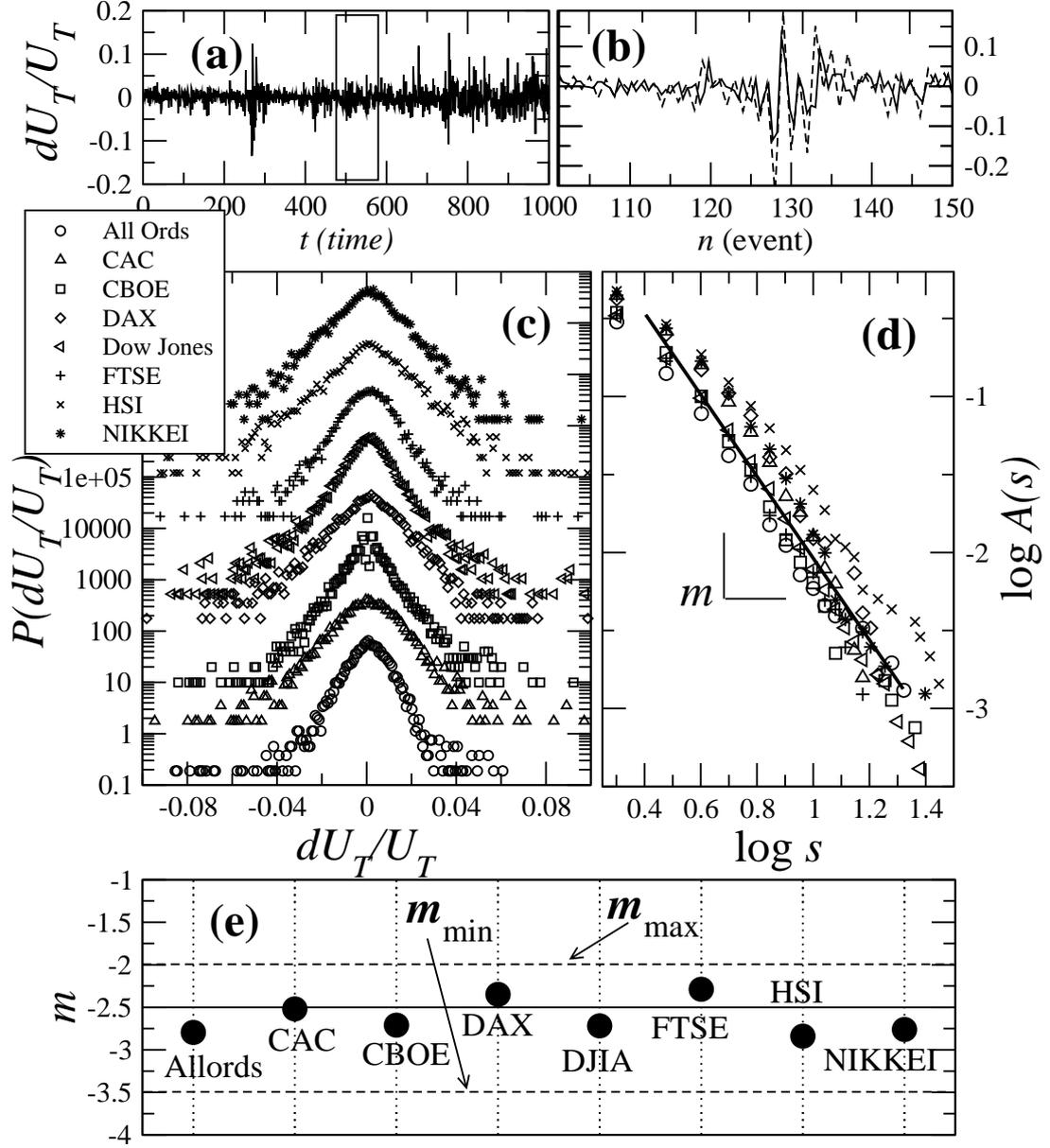}}
\vspace*{8pt}
\caption{\protect
    {\bf (a)} Time evolution of the logarithmic returns of the DJIA index 
(partial). The box is zoomed in {\bf (b)} to emphasize the contrast 
between real data (solid line) throughout time $t$ and the succession of
events (dashed line) where the variation changes sign.
{\bf (c)} Probability density functions for some important financial 
indices, including interest rate options (CBOE).
{\bf (d)} Distribution of avalanche sizes detected throughout the evolution 
of each financial index showing critical behavior with 
{\bf (e)} an exponent $m$ approximately invariant and similar to the
one obtained for branching processes and to our model (see tex).
Solid line in (e) indicates the theoretical value obtained for
$\gamma=3$ (see Eq.~(\ref{eq:avalanchefinalacum})), 
while dashed lines indicate the
bounding values of $m$ for $2 < \gamma < 3$ 
(see Sec.~\ref{sec:distributionbounderies}).}
\label{fig5} 
\end{figure}
%%%%%%%%%%%%%%%%%%%%%%%%%%%%%%%%%%%%%%%%%%%%%%%%%%%%%%%%%%%%%%%%%%%%%%%%%%
which connects the exponent $m$
characterizing dynamical events (avalanches) with $\gamma$, the
exponent characterizing the structure (degree distribution) of 
the underlying economical network.

Looking again to Fig.~\ref{fig3}b, with the help of
Eq.~(\ref{eq:avalanchefinalacum}) and borrowing from the 
literature\cite{barabasi} the values of $\gamma$ of empirical 
networks, typically in the range $[2.1,2.7]$,
one concludes that the size distribution exponent should take typical
values $m\in [2.15,3.05]$ which agrees with the results from our model,
as shown not only in Fig.~\ref{fig3}b but also in Fig.~\ref{fig5}
which presents data from several stock market indexes.

The time-series of the logarithmic returns shown in
Fig.~\ref{fig5}a must first be mapped in an series of events
as shown in Fig.~\ref{fig5}b. 
One event is defined as a (typically small) set of successive instants 
in the original time-series having the same derivative sign, either 
positive (monotonically increasing values) or negative
(monotonically decreasing values). 
Each time the derivative changes sign a new event starts. 
In the continuous limit, events would correspond
to the instants in the time-series with vanishing 
first-derivative.

The non-Gaussian distribution of the logarithmic returns 
(Fig.~\ref{fig5}c) were extracted from the logarithm
returns of the original series of each index, similarly to
what is done in Ref.~\cite{Kiyono2006}. The characteristic heavy 
tails observed by Kiyono and co-workers 
are observed for short time
lag (hours or smaller), wheres in Fig.~\ref{fig5}c the daily 
closure values are considered.
The power-law behavior of the avalanche size 
(Fig.~\ref{fig5}d) is indeed similar to the simulated results
in Fig.~\ref{fig3}b.

Figure \ref{fig5}e shows the exponent $m$ in
Eq.(\ref{eq:avalanchefinalacum2}) for each one of the stock market
indexes. All of them take values around the average theoretical prediction
$m=5/2$.

Further, all values lay between to bounds, 
\begin{equation}
m_{min}\equiv 2 < m < \tfrac{7}{2} \equiv m_{max}, 
\label{mmin_mmax}
\end{equation}
as we explain in Section \ref{sec:distributionbounderies}. 

These two limits play a major role in the description of the critical
behavior. Further, since a return distribution can be bounded, it is possible 
to measure the risk of a wrong model choice.  
%%%%%%%%%%%%%%%%%ISTO TB NAO%%%%%%%%%%%%%%%%%%%%%%%%%%%%%%%%%%%
%and
%consequently whatever model we use to calculate risk over financial 
%returns we can measure the risk of that model being incorrect.
%%%%%%%%%%%%%%%%%%%%%%%%%%%%%%%%%%%%%%%%%%%%%%%%%%%%%%%%%%%%%%%%%  
That kind of risk is called ``model risk''. 
%Since it is possible to know the boundaries $m_{min}$ and $m_{max}$, 
%the usage of any risk model, based on Gaussian curves or not, 

%Adopting the limits derived here, it is possible to establish
%upper and lower boundaries for the probability of observing avalanches
%of a given size. In other words, the boundaries $m_{min}$ and $m_{max}$
%are able to account for the risk associated to one other model
%being used.}
%%%%%%%%%%% ISTO NAO SE ENTENDE... %%%%%%%%%%%%%%%%%%%%%%%%%%%%%%
%to every single situation where there 
%is a need for quantifying financial risk 
%can be regard as absurd because risk measures are needed for investment 
%purposes and 
%this would mean that every investment would have the same risk measure. 
%But, with 
%these limits we can quantify the amount in which the risk measure of 
%the investment can be wrong, a second measure of risk that can become 
%very important in regulatory issues.

The result in Eq.~(\ref{eq:avalanchefinalacum}) relates 
the microscopic economic relations -- trades among agents -- with 
the form of the heavy-tailed return distributions.
This relation emerges from the interplay between the distribution of 
the returns, one macroscopic quantity chosen as the overall product
$U_T$, and the topology of the social network.
The economic connection was defined as one exchange of labor, in one
of its forms. No additional assumptions were taken. Since we never 
made any conjectures regarding the specific type of economic interaction.
Equation (\ref{eq:avalanchefinalacum}) should hold for all types of 
trades and macroscopic observables of the economical product in one 
economic network, since we instantiate economic connections on the more 
abstract level possible. 

All empirical indices are sampled daily but in different time periods.
For FTSE $6498$ days in London stock market were considered,
starting on April 2nd 1984 and ending on December 18th 2009.
For DJIA $20395$ days in New York stock market were considered,
starting on October 1st and ending on December 18th 2009.
For DAX $4815$ days in Frankfurt stock market were considered,
starting on November 26th 1990 and ending on December 18th 2009.
For CAC $5003$ days in Paris stock market were considered,
starting on March 5th 1990 and ending on December 18th 2009.
For ALLORDS $6555$ days in Australian stock market were considered,
starting on August 3rd 1984 and ending on June 30th 2010.
For HSI $5701$ days were considered in Hong Kong stock market, 
starting on December 31st 1986 and ending on December 18th 2009.
For NIKKEI $6386$ days were analyzed in Tokyo stock market, 
starting on January 4th 1984 and ending on December 18th 2009.
For CBOE IR10Y $12116$ days in Chicago derivative market,
starting on January 2nd 1962 and ending on June 30th 2010.
%%%%%%%%%%%%%%%%%%%%%%%%%%%%%%%%%%%%%%%%%%%%%%%%%%%%%%%%%%%%%%%%%%%%
\begin{figure}[t]
\includegraphics[width=14.0cm,angle=0]{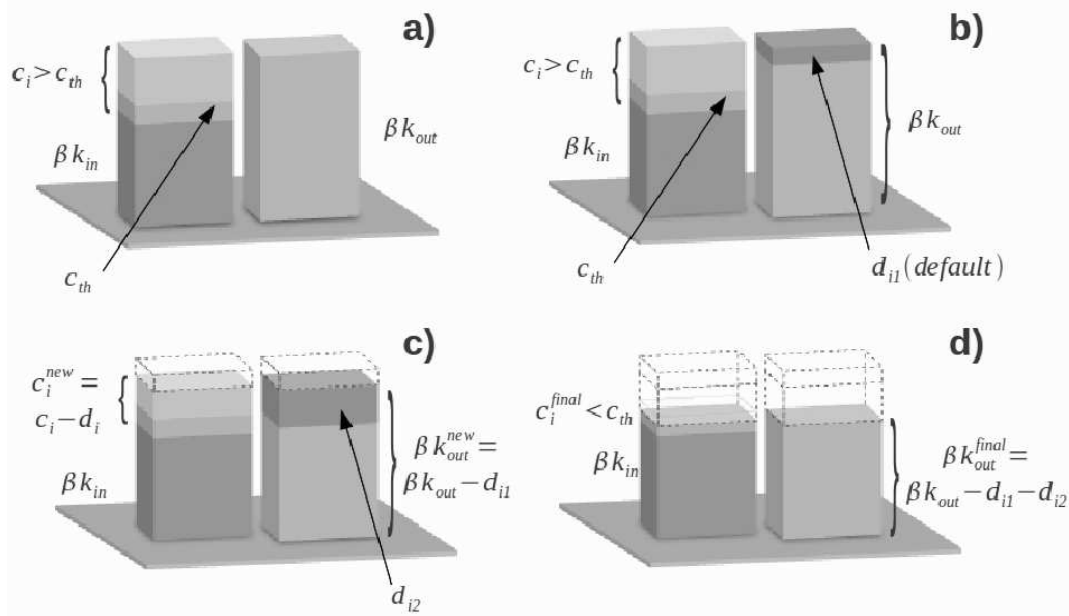}
 \caption{\protect 
         Sketch of banks system resilience. 
         {\bf (a)} Bank balance with no defaults, having a capital 
                   $c_i=\beta(k_{out}-k_{in})>c_{th}$ above the minimum capital 
                   level $c_{th}$. 
         {\bf (b)} After client default $d_{i1}$, the number of outgoing 
                   connections will be reduced, 
         {\bf (c)} reducing the number of shareholder connections to 
                   $\beta k_{out}^{new}=\beta k_{out}-d_{i1}$ and yielding a new 
                   capital investment $c_i^{new}=c_i-d_{i}$. 
         If client default persists, at a new level $d_{i2}$, 
         {\bf (d)} the capital may drop bellow minimum level 
                   $c_i^{final}<c_{th}$, leading to bankruptcy (see text).}
\label{fig7}
\end{figure}
%%%%%%%%%%%%%%%%%%%%%%%%%%%%%%%%%%%%%%%%%%%%%%%%%%%%%%%%%%%%%%%%%%%%

%%%%%%%%%%%%%%%%%%%%%%%%%%%%%%%%%%%%%%%%%%%%%%%%%%%%%%%%%%%%%%%%%%%%%%%%%%%%%
\section{Consequences of heavy-tails in financial stability}
\label{sec:stability}

In this section we apply our model to investigate the consequences of
recent directives in the banking system, which are totally based in the 
Merton-Vasicek model for credit portfolio risk\cite{MertonVasicek}, and 
simulate it in a P1,P2 and P3 context.

Banks also follow common rules, settled under 
general agreement. One very important rule is to fix a minimum fraction 
of invested money as money belonging to the shareholders, i.e.~to the 
bank itself, and not to the depositors. Such money is called minimum
capital level and the rules to regulate it establish resilience for 
the bank system. 
This minimum capital level is a direct financial 
translation of postulate P3, 
because it imposes a minimum threshold to banks that 
what to stay in business.
 
Since a bank could, in principle, collect an arbitrary amount 
of money from its depositors and invest it -- or loose it -- each country 
created its own rules for the minimum amount of capital, 
expressed as a percentage of the total investment. In 
1998, a group of central bank governors called the Basel Committee on Banking 
Regulation unified these rules to all banks, intending to protect the global 
banking system\cite{baselhist}. % and to eliminate competitive inequality
Among other, these rules impose a minimum capital level at 8\% without any 
empirical reason\cite{basel1}.
In the second version of the accord\cite{basel2}, the Value-at-Risk 
paradigm\cite{Holton} was adopted to calculate the risk weights.
 
The social pressure over the Committee to tighten the minimum capital 
rules became very strong after the 2008 financial turmoil, 
which lead to an unexpected avalanche of bank insolvencies all around the 
globe, questioning the ability of regulators to make effective bank system 
protection rules. Thus, in 2010 they issue the third version, Basel 
III\cite{basel3}, to improve bank system resilience by 
raising the levels of capital.
At the same time, these directives were subjected to significant 
criticism by the academic community\cite{Danielsson2001} arguing how they 
``even created the potential for new sources of instability''.

In this section we address this critic to nowadays bank regulation rules
by showing that %in a P1, P2 and P3 context
the bank system resilience does not 
necessarily improve with such raise of the capital levels.
To this end we instantiate our model for generic economic networks 
into the banking network and enable to vary the threshold $c_{th}$.

The agents represent now banks, which provide production 
of credit (money) to their neighboring banks, 
through an integer number $k_{out,i}$ 
of outgoing connections, using its own capital $c_i$ and gets credit 
through other $k_{in,i}$ incoming connections from agents %other agents, 
which make deposits in bank $i$.
We call the first (outgoing) connections the creditor connections and 
the second (incoming) connections the debtor connections.
See Fig.~\ref{fig7}.
The chain of collapses are now taken as chains of bankruptcies.

The bankruptcy of a bank $i$ occurs when the number 
of destructed creditor connections becomes such that the number 
of shareholder debtor connections drops below a regulatory 
minimum level, i.e.~${c}_{th}>{c}_i$, leading to the removal of all 
banks debtor connections of bank $i$, setting $k_{in}=1$, implying
an update in bank $i$ and its neighbors $j$ according to
Eqs.~(\ref{eq:avalanche}).

Once again,
having an agent that bankrupts, the natural question that follows
is what is the probability for that bankruptcy to trigger an avalanche of
bankruptcies (financial global crisis).
In the situation above of infinite leverage, when one agent is bankrupted
the average number of neighbors that are induced to bankrupt is smaller 
than one. In the situation of zero leverage this average is larger than one.
Since the financial system remains “trapped” between two phases, 
the average number of bankrupted agents that follow a bank 
bankruptcy in a chain reaction must be equal to one. And this makes all the 
difference in terms of banking system stability, especially when we study 
the magnitude of the avalanche.
%%%%%%%%%%%%%%%%%%%%%%%%%%%%%%%%%%%%%%%%%%%%%%%%%%%%%%%%%%%%%%%%%%%%%%%%%%
\begin{figure}[t]
\begin{center}
\includegraphics[width=12.0cm,angle=0]{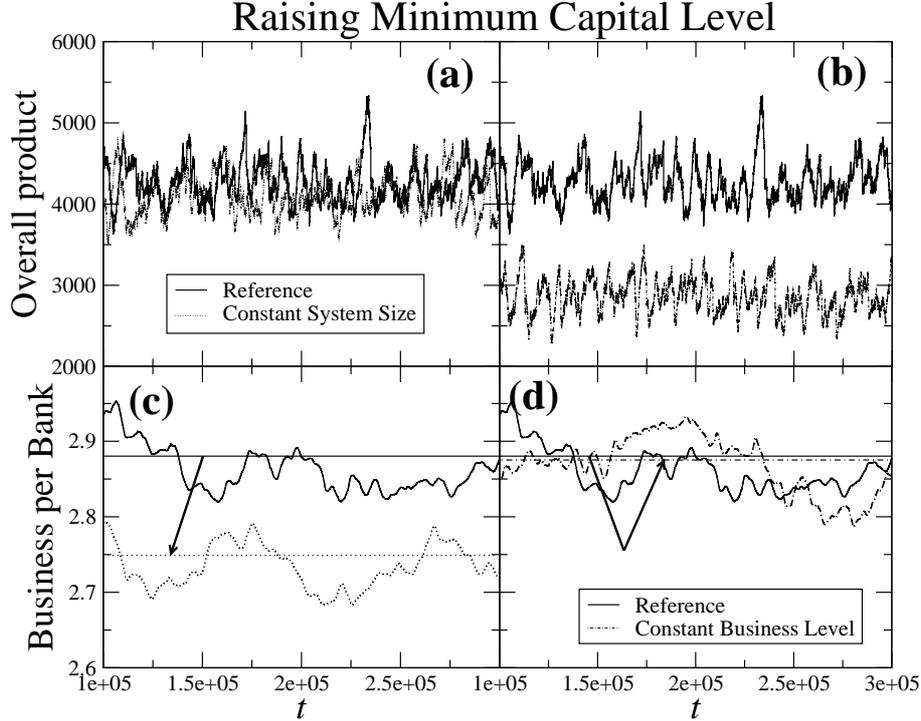}
\end{center}
\caption{\protect          
         Considering the overall product $U_T$ (units of $W$) 
         of a reference minimum 
         capital level ${c}_{th}=-0.71$ (solid line), 
         there are two different processes 
         for achieving quasi-stationary states when minimum capital level 
         is increased up to ${c}_{th}=-0.69$
         {\bf (a)} by keeping the number of agents constant 
         at L=2000 (dotted line) or 
         {\bf (b)} by maintaining the business level constant 
         at $\Omega=2.88$ in units of $W$ (dotted line), which leads to 
         a decrease of $L=1500$.           
         {\bf (c)} The average business level $\Omega$ per bank 
         (Eq.~(\ref{eq8})) decreases when raising the
         minimum capital level and keeping the number of agents constant.
         {\bf (d)} Assuming that banks do not want to see their business level 
         dropping, it is natural to expect that they will choose a process 
         with constant business level to face an increase of the minimum 
         capital level.}
\label{fig8}
\end{figure}
%%%%%%%%%%%%%%%%%%%%%%%%%%%%%%%%%%%%%%%%%%%%%%%%%%%%%%%%%%%%%%%%%%%%%%%%%%

Summing up the product due to the creditor connections in the network 
one gets the overall product $U_T$ in Eq.~(\ref{eq:totalenergy}).
Figure \ref{fig8}a and \ref{fig8}d
shows the evolution of overall products for two different ways
of raising minimum capital levels. 
One time step corresponds to one new
trade connection.
Black solid line is the reference state.
In Fig.~\ref{fig8}a minimum capital level ${c}_{th}$
is raised keeping the size of the system constant (green dashed line),
while Fig.~\ref{fig8}b shows what happens when the same raise of
minimum capital level is accompanied by a decreasing of the system size.

Keeping constant the number of agents that trade within the system, 
means for each agent to maintain the same neighborhood as previously, 
before the raise of minimum capital level. 
We call the system incorporating this neighborhood, the operating 
neighborhood and its size is given by the number $L$ of agents.
The raise of minimum
capital level can however occur by maintaining other properties
constant. Namely, the average business level per bank, defined as 
the moving average in time of $U_T$ per bank:
\begin{equation}
\Omega=\frac{1}{L}\frac{1}{T_S}\int_{t}^{t+T_S} U_T(x)dx
\label{eq8}
\end{equation}
where $T_S$ is a sufficiently large period for taking time averages
and the $dt$ in the integrand is given by the time step of our
simulation. 
Similar quantities are taken in Economics as indicators of 
individual average standards of living\cite{tina}.
Roughly the time derivative of $\Omega$ gives the overall product 
uniformly distributed by all agents in the network.
Figures \ref{fig8}c and \ref{fig8}d shows the business level per bank
corresponding to the overall product in Figs.~\ref{fig8}a and \ref{fig8}b
respectively. In particular, while in Fig.~\ref{fig8}c the business
level decreases significantly with the raise of ${c}_{th}$, in 
Fig.~\ref{fig8}d
the business level is approximately constant, with
the smaller size of the operating neighborhood normalizing the
overall product. See
horizontal lines, showing average values for business level.

Figure \ref{fig8} shows that keeping constant the size $L$ 
of the operating neighborhood while
raising the minimum capital level induces a decrease of the business 
level. 
Differently, if the business level is kept constant, the size of the 
operating neighborhood shrinks.
One should stress that in real financial systems, to keep the same 
business level, agents merge which indeed induces the
network to shrink.
%%%%%%%%%%%%%%%%%%%%%%%%%%%%%%%%%%%%%%%%%%%%%%%%%%%%%%%%%%%%%%%%%%%%%%%%%
\begin{figure}
\begin{center}
\includegraphics[width=0.9\textwidth,angle=0]{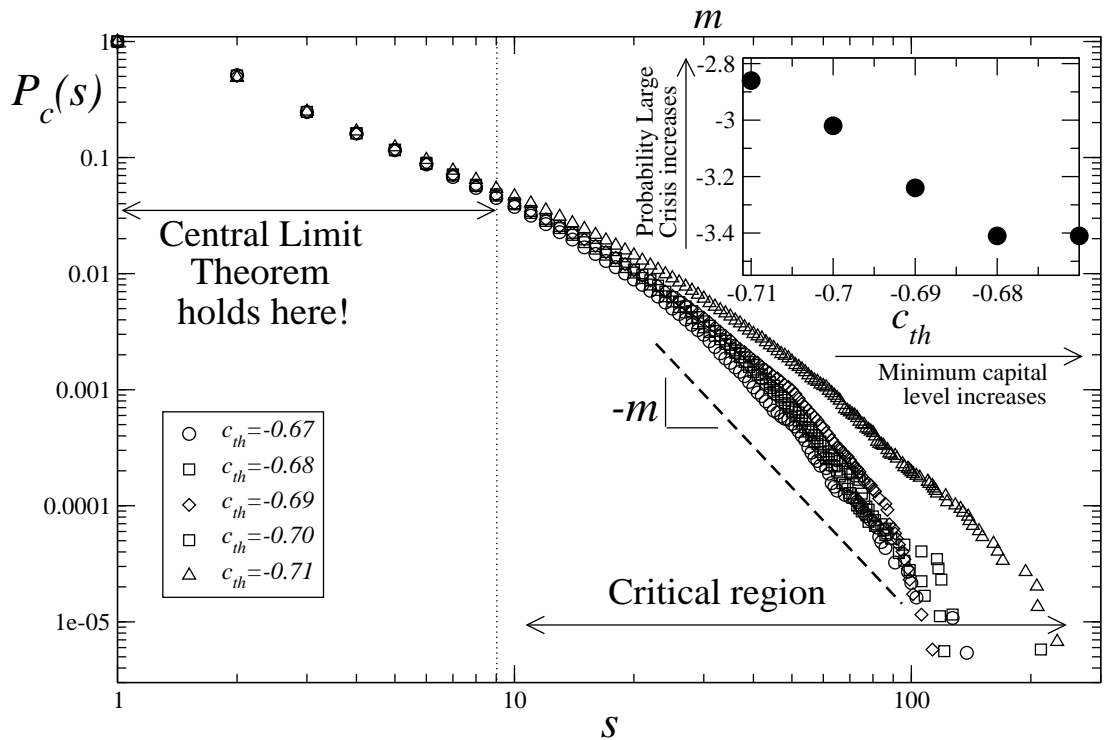}
\end{center}
\caption{\protect 
        Keeping the same operating neighborhood for each agent ($L=2000$),
        different minimum capital levels yield different avalanche (crisis) 
        size distributions. 
        For small sizes the Central Limit Theorem holds and different
        distribution match with each other.
        Differently, large size avalanches occur at critical states of the
        system, and therefore the distributions deviate from  each other.
        In the inset one shows the exponent obtained for the critical region
        for each scenario of minimum capital level: increasing 
        minimum capital level decreases the probability for a large avalanche
        to occur, which supports the intentions of Basel III accords. However,
        such behavior is only guaranteed under the assumption of keeping
        the same operating neighborhood $L$ for the minimum level raise,
        a scenario which is not natural in a trading system of economic
        agents (see text and Fig.~\ref{fig10}).}
\label{fig9}
\end{figure}
%%%%%%%%%%%%%%%%%%%%%%%%%%%%%%%%%%%%%%%%%%%%%%%%%%%%%%%%%%%%%%%%%%%%%%%%%

Next we investigate the first situation, i.e.~we consider a raise of the 
minimum capital level with constant $L$.  
To this end we compute the fraction $P_c(s)$ 
of avalanches of size larger than $s$, yielding the cumulative size 
distribution of avalanches. Numerically $P_c(s)$ for a given $s$ is
obtained by counting at each 
iteration the total number of debtor connections involved in each chain of 
bankruptcies, choosing those chains whose total number is larger than $s$.

Figure \ref{fig9} shows the cumulative size distributions of 
avalanches in our model for different minimum capital levels, 
keeping the number of agents constant ($L=2000$).
For small avalanche sizes, the Central Limit Theorem 
holds\cite{Mantegna_Stanley1995}
and thus all size distributions match independently of the minimum
capital level.
For large enough avalanches (`critical region'), the size distributions
deviate from each other, showing a power-law tail $P_c(s)\sim s^{-m}$ 
with an exponent $m$ depending on the minimum capital level ${c}_{th}$.

\begin{figure}
\includegraphics[width=8.0cm,angle=0]{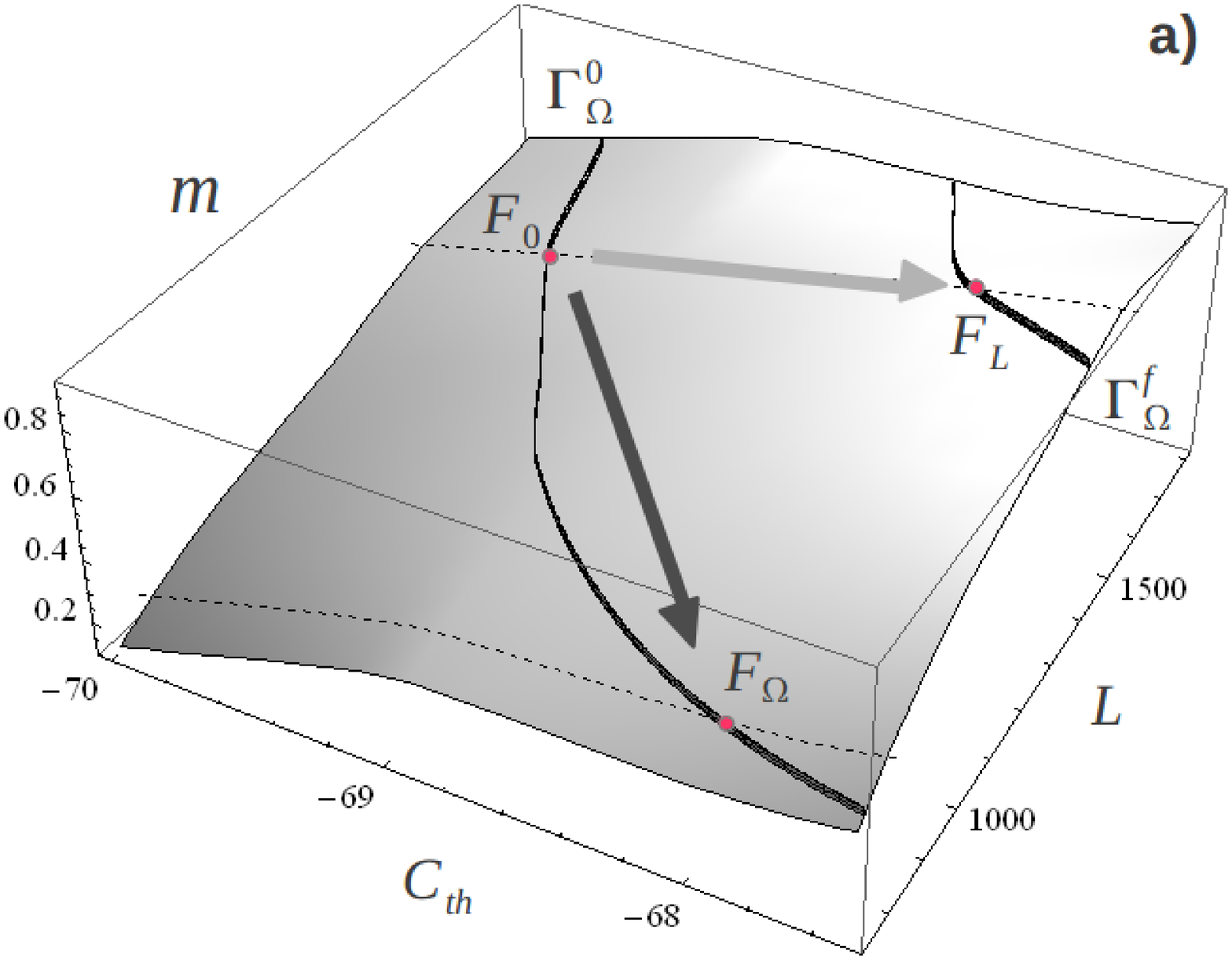}
\includegraphics[width=8.0cm,angle=0]{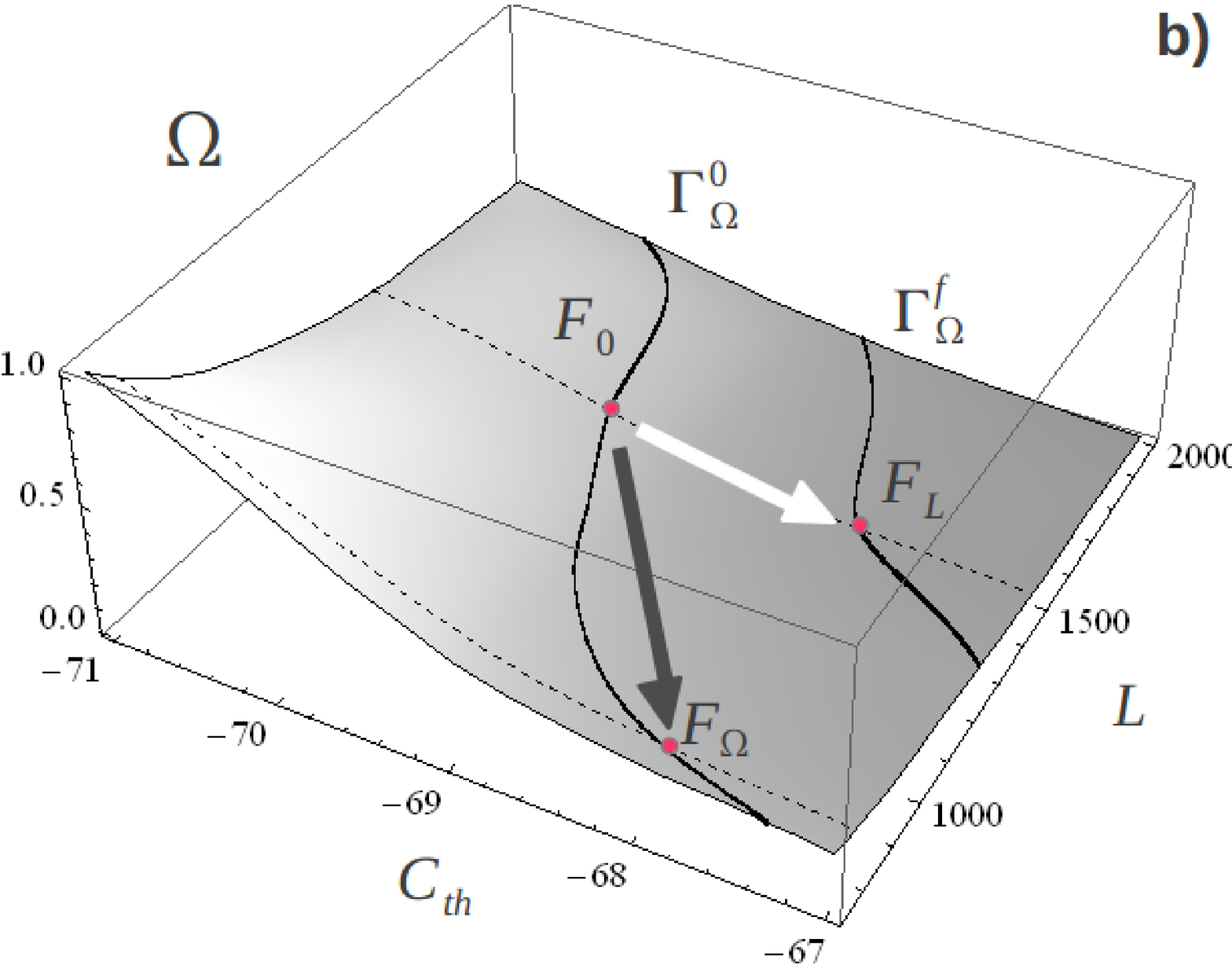}
\caption{\protect 
         {\bf (a)} Normalized critical exponent $\bar{m}$ 
         as a function of the minimum capital level 
         ${c}_{th}$ and the size $L$ of the operating neighborhood.
         For an initial financial state $F_0$ an increase of the minimum 
         capital level to state $F_L$ means to follow an isoline of
         constant $L$. This leads to a larger $m$ value.
         %, smaller
         %probability for a larger avalanche to occur.
         A path to a different state $F_{\Omega}$ having the same minimum 
         capital level raise can be reached, following an isoline of
         constant business $\Omega$.
         In this case a smaller value for $m$ is found.
         %, meaning a larger
         %probability for large avalanches to occur.
         As shown in plot
         {\bf (b)} while the business $\Omega$ remains constant in the
         latter case, for the former case (constant $L$) it decreases.
         Assuming that agents will try to optimize their gains, it is
         more naturally to expect for them to follow isolines of constant
         business level which leads to the increase of probability for
         large crisis to occur (see text).}
\label{fig10}
 \end{figure}

As one sees in the inset of 
Fig.~\ref{fig9} the exponent increases in absolute value for larger minimum 
capital levels, which prevents large avalanches to occur. Though, such scenario 
occurs only when the size (number of agents) of the financial subsystem 
where the agent makes its trades is kept constant. 
For a scenario where the size of the operating neighborhood is adjusted to 
maintain the business level constant Eq.~(\ref{eq8}) the situation is 
different as shown next, in Fig.~\ref{fig10}.
%%%%%%%%%%%%%%%%%%%%%%%%%%%%%%%%%%%%%%%%%%%%%%%%%%%%%%%%%%%%%%%%%%%%%%%%%
%%%%%%%%%%%%%%%%%%%%%%%%%%%%%%%%%%%%%%%%%%%%%%%%%%%%%%%%%%%%%%%%%%%%%%%%%

Figure \ref{fig10}a and \ref{fig10}b
show the critical exponent $m$ and the business level respectively,
as a function of the minimum capital level  ${c}_{th}$ and the operating 
neighborhood size $L$. 
For easy comparison, both quantities are normalized in the unit 
interval of accessible values.

Roughly, the critical exponent shows a tendency to increase
with both the minimum capital level and the operating neighborhood size,
while with the business level the opposite occurs.
Considering a reference state $F_0$ with
$c_{th,0}$, $L_0$ and $\Omega_0$ there is one isoline of constant system size, 
$\Gamma_{L}^0$ and another of constant business level $\Gamma_{\Omega}^0$, 
both crossing at $F_0$. 

From Fig.~\ref{fig10}a, one sees that assuming a transition from state 
$F_0$ to a state with larger minimum 
capital level keeping the system size constant, one follows 
the isoline $\Gamma_L^0$, arriving to the new state $F_L$.
This state has a larger value of the critical exponent $m$, which means 
a lower probability for large avalanches to occur, as explained above. 
However in such situation the new business level $\Omega_f < \Omega_0$ 
is smaller than the previous one, as clearly illustrated in Fig.~\ref{fig10}b.

On the contrary, if we assume the transition to the higher minimum capital 
level occurring at constant business level, i.e.~along the isoline 
$\Gamma_{\Omega}^0$ one arrives to a state $F_{\Omega}$ for which the critical 
exponent is approximately the same, with a larger probability for large 
avalanches to occur. 

From economical and financial reasoning, one assumes typically
that independently of external directives, under unfavorable situations,
economical and financial agents try, at least, to maintain their business 
level.
Such behavior from agents leads to a situation which
contradicts the expectations in Basel accords and raises the 
question if such regulation will indeed prevent a larger avalanches 
to occur again in the future.

%%%%%%%%%%%%%%%%%%%%%%%%%%%%%%%%%%%%%%%%%%%%%%%%%%%%%%%%%%%%%%%%%%%%%%%
\section{Conclusions}
\label{sec:conclusion}            

The emergence of heavy-tails on distributions associated with
economic and financial phenomena has a direct link with the underlying 
mechanisms of interaction. Looking from the perspective of the distribution 
we can deduce the type of mechanisms and from the type of mechanisms we can 
explain how the heavy-tails occur.

In this chapter we showed the basic postulates describing economic 
interactions that underlying the emergence of heavy tails in economy:
(i) all individuals tend to establish trades, 
(ii) different individuals have different attractiveness to get into new trades,
(iii) all individual have finite leverage.

We incorporated such postulates in a minimal model to demonstrate why a 
economic system is a system in a critical state. Further, we
showed that the geometrical structure composed by the trade connections
in the entire economic network has constraints that reflect a finite
range for the values of exponents describing heavy-tails. These boundary
values are able to account for improved risk measures.

A specific application of our model for critical behavior among economic 
agents was also addressed, in the context of global banking regulation,
showing that the actual tendence and principles for establishing 
banking rules at the level of individuals bank entities can have been 
driven on the opposite direction of its goals, since they ignore the 
complex nature of the economic network that is revealed by 
the heavy-tailed distributions of the variables that characterize the system.

Two notes are due here. 
First, in practice, capital is a consumption of 
wealth from agents called shareholders and could be represented by the 
amount $k_{in}$. The reason why we do not do it is because banking 
regulation separate shareholders from depositors and to map the two 
approaches we need to make the same segregation.
Second, the difference 
between credit connections and debtor connections is not equal to 
the difference between assets and liabilities, it only represents it. That 
is, the break of the debtor connections on the agents that connect 
with the bank represents a destruction of a creditor connection, 
an asset. 
The main difference resides on the fact that we are not interested in fixed 
or non-performing assets. A default represents a destruction of a creditor
connection in opposition to bank accounting where the loan becomes a 
depreciating stock.
 
%%%%%%%%%%%%%%%%%%%%%%%%%%%%%%%%%%%%%
\section*{Acknowledgments}
The authors thank Nuno Ara\'ujo for helpful discussions and
also PEst-OE/FIS/UI0618/2011 for partial financial support.
PGL thanks 
{\it Funda\c{c}\~ao para a Ci\^encia e a Tecnologia – Ci\^encia 2007} for
financial support.

%%%%%%%%%%%%%%%%%%%%%%%%%%%%%%%%%%%%%%%%%%%%%%%%%%%%%%%%%%%%%
%\section*{List of Symbols\hfill} 
%\addcontentsline{toc}{section}{List of Symbols}
%\listofsymbols

%%%%%%%%%%%
%\begin{thebibliography}{00}
\bibliographystyle{plain}
\bibliography{heavytailscompsystems_cl.bib}
%\end{thebibliography}

%%%%%%%%%%%%%%%%%%%%%%%%%%%%%%%%%%%%
%%%%%%%%%%%%%%%%%%%%%%%%%%%%%%%%%%%%
\end{document}